\documentclass{article}
\pdfoutput=1
\usepackage[margin=2cm]{geometry}
\usepackage{graphicx,bm,amsmath,amssymb,latexsym,psfrag,epsfig,euscript, multirow,color,verbatim,cancel}
\usepackage{mathrsfs,tabularx}
\usepackage{float}
\usepackage{subfigure}
\usepackage{cite}
\usepackage{color}
\definecolor{darkgreen}{rgb}{0,0.5,0}

\allowdisplaybreaks
\begin{document}
\newcolumntype{C}{>{\centering\arraybackslash}X}
\newcommand{\PL}{\mathcal{P}_{\mathrm L}}
\newcommand{\PR}{\mathcal{P}_{\mathrm R}}
\newcommand{\gs}{g_{\mathrm s}}
\newcommand{\be}{\begin{equation}}
\newcommand{\ee}{\end{equation}}
\newcommand{\beq}{\begin{equation}}
\newcommand{\eeq}{\end{equation}}
\newcommand{\bean}{\begin{eqnarray*}}
\newcommand{\eean}{\end{eqnarray*}}
\newcommand{\bea}{\begin{eqnarray}}
\newcommand{\eea}{\end{eqnarray}}
\newcommand{\MPl}{M_{\mathrm{Pl}}}
\newcommand{\Td}{T_{\mathrm{d}}}
\newcommand{\ptl}{\partial}
\newcommand{\benum}{\begin{enumerate}}
\newcommand{\eenum}{\end{enumerate}}
\newcommand{\bi}{\begin{itemize}}
\newcommand{\ei}{\end{itemize}}
\newcommand{\tb}{\tilde{B}}
\newcommand{\tw}{\tilde{W}}
\newcommand{\tg}{\tilde{g}}
\newcommand{\tH}{\tilde{H}}
\newcommand{\bs}[1]{\textcolor{darkgreen}{[BS: #1]}}
\newcommand{\yc}[1]{\textcolor{blue}{[YC: #1]}}
\begin{titlepage}

\begin{flushright}
UMD-PP-014-014\\
NSF-KITP-14-157
\end{flushright}

\vspace{0.2cm}
\begin{center}
\Large\bf
Probing Baryogenesis with Displaced Vertices at the LHC
\end{center}

\vspace{0.2cm}
\begin{center}
{
Yanou Cui$\,^{a,b}$\footnote{E-mail:~cuiyo@umd.edu} and Brian Shuve$\,^{b}$\footnote{E-mail:~bshuve@perimeterinstitute.ca}\\
\vspace{15pt}
$^{a}$Department of Physics, University of Maryland, College Park, MD 20742, USA.\\
 $^{b}$ Perimeter Institute for Theoretical Physics, Waterloo, Ontario, Canada N2L 2Y5.
}
\end{center} 

\vspace{0.2cm}
\begin{abstract}\vspace{0.2cm}
\noindent 

\setcounter{footnote}{0}

The generation of the asymmetric cosmic baryon abundance requires a departure from thermal equilibrium in the early universe. In a large class of baryogenesis models, the baryon asymmetry results from the out-of-equilibrium decay of a new, massive particle. We highlight that in the interesting scenario where this particle has a weak scale mass, this out-of-equilibrium condition requires a proper decay length larger than $O(1)$ mm. Such new fields are within reach of the LHC, at which they can be pair produced leaving a distinctive, displaced-vertex signature. This scenario is realized in the recently proposed mechanism of baryogenesis where the baryon asymmetry is produced through the freeze-out and subsequent decay of a meta-stable weakly interacting massive particle (``WIMP baryogenesis''). In analogy to missing energy searches for WIMP dark matter, the LHC is an excellent probe of these new long-lived particles responsible for baryogenesis via the low-background displaced vertex channel. In our paper, we estimate the limits on simplified models inspired by WIMP baryogenesis from two of the most sensitive collider searches by CMS and ATLAS with 8 TeV LHC data. We also estimate the LHC reach at 13 TeV using current strategies, and demonstrate that improvements of one to two orders of magnitude in cross-section limits can be achieved by requiring two displaced vertices, with further improvement possible with lower kinematic thresholds. For meta-stable WIMPs produced through electroweak interactions, the high luminosity LHC is sensitive to masses up to 2.5 TeV for lifetimes around 1 cm, while for singlets pair-produced through the off-shell-Higgs portal, the LHC is sensitive to production cross sections of $O(50)$ ab for benchmark masses around 150 GeV. Our analysis and proposals also generally apply to displaced vertex signatures from other new physics such as hidden valley models, twin Higgs models and displaced supersymmetry.

\end{abstract}
\vfil

\end{titlepage}

\tableofcontents

\section{Introduction}
Despite our familiarity with the physics of baryonic matter, the origin of the cosmic asymmetric abundance of baryons \cite{Ade:2013zuv} remains one of the most prominent questions that demands new physics beyond the Standard Model (SM), and is as puzzling as the nature and interactions of dark matter (DM). Various  baryogenesis mechanisms have been proposed and studied over the years, such as leptogenesis \cite{Fukugita:1986hr}, the Affleck-Dine mechanism \cite{Affleck:1984fy}, and electroweak baryogenesis \cite{Kuzmin:1985mm,Shaposhnikov:1986jp,Shaposhnikov:1987tw}; of these, only electroweak baryogenesis has a necessary link to the weak-scale, while the others can operate at much higher scales, making them challenging to probe experimentally. Although some of these baryogenesis scenarios may be tested indirectly via electric dipole moments \cite{Pospelov:2005pr}, neutron-antineutron oscillation \cite{Babu:2013jba, Babu:2013yww}, and neutrino-less double beta decay \cite{Vergados:2012xy,Rodejohann:2012xd}, the actual baryogenesis mechanism cannot typically be reproduced  at current laboratory experiments such as the Large Hadron Collider (LHC). Even in electroweak baryogenesis, the generation of the asymmetry occurs during the electroweak phase transition, which is irreproducible at current experiments. This is in contrast with the situation for dark matter: when the dark matter abundance is established through thermal freeze-out such as in the compelling case of weakly interacting massive particle (WIMP) DM (known as the ``WIMP miracle''), the LHC \cite{Petriello:2008pu,Gershtein:2008bf,Cao:2009uw,Beltran:2010ww,Goodman:2010yf,Bai:2010hh} and DM detection experiments \cite{Goodman:1984dc} can directly probe the interactions responsible for freeze-out due to crossing symmetry. In addition to better testability, baryogenesis based on new weak-scale particles is also theoretically appealing as it can naturally connect new physics addressing the weak-scale hierarchy problem with the dynamics responsible for generating the baryon asymmetry in our universe, which is a profound cosmological problem. This is analogous to the appeal of thermal WIMP DM, where DM candidates can emerge from motivated models such as supersymmetry. A few examples of low-scale baryogenesis models that generate a baryon asymmetry via the decays of new weak-scale states have been shown to have direct testability at colliders \cite{Blanchet:2009bu,Fong:2013gaa, An:2013axa,AristizabalSierra:2010mv,Deppisch:2013jxa,Sierra:2013kba}.

In this paper, we highlight an interesting observation: for a generic possibility that baryogenesis is triggered by the decay of a new weak-scale particle $\chi$, the Sakharov condition  requiring a departure from equilibrium \cite{Sakharov:1967dj} is  robustly satisfied when the particle lifetime exceeds the Hubble time $H^{-1}$ at temperatures $T\sim M_\chi$ \cite{Kolb:1990vq}. This in turn implies that the proper decay length of the new state satisfies $c\tau\gtrsim H^{-1} |_{T=M\chi}\approx O(1)$ mm for $M_\chi \sim O(100)$ GeV. In this scenario, the washout of the baryon asymmetry from inverse decays is always suppressed, leading to the effective preservation of the baryon asymmetry from $\chi$ decays\footnote{Baryogenesis can also be viable in the ``strong washout'' regime, $\Gamma \gtrsim H(M_\chi)$, provided the asymmetry generated after asymmetry-destroying interactions have frozen out is sufficiently large. This is the case in models such as leptogenesis from right-handed neutrino decays \cite{Buchmuller:2004nz} and WIMPy baryogenesis (baryogenesis via DM annihilation) \cite{Cui:2011ab}, and the baryon asymmetry is sensitive to the specific details of washout freeze-out.}. Interestingly, this cosmological condition on the proper decay length of $\chi$ has implications for its collider phenomenology: {\bf if $\chi$ can be produced at a collider, it travels a macroscopic distance $\gtrsim1$ mm before decaying at a displaced vertex (DV)}\footnote{If the new particle is \emph{very} long-lived and travels $\gg O(1)$ m before decaying, it appears as missing energy and is covered by conventional dark matter searches, which have relatively lower sensitivity due to the reduction in signal rate from having to tag visible objects (such as initial state radiation), and larger SM backgrounds.}.

DVs require dedicated collider search strategies, as  most conventional collider searches place quality cuts on  tracks, excluding those with sufficiently large impact parameters: less than $O$(cm) at LEP, and less than $O$(mm) at the Tevatron and LHC detectors. Due to the difficult and specialized nature of reconstructing displaced tracks and vertices, and because displaced vertices have only relatively recently been realized as generically motivated by beyond-the-SM theories, DV searches have not received as much attention as other searches based on prompt decays or missing transverse energy, although they have drawn rising interest and efforts in the past few years with much room still remaining for further development. The particular promise and appeal of DV searches lies in their exceptional sensitivity:~with essentially no irreducible SM backgrounds, DV searches can probe tiny signal cross sections as long as the signals can be efficiently triggered upon and reconstructed. As we discuss later, current search strategies do not always fully exploit the signal characteristics, and some searches may consequently have large backgrounds at 13 TeV with high luminosity, while the signals can have low efficiency of passing triggers, reducing the sensitivity at 13 TeV.  Existing theoretical motivations for DV searches include gauge-mediated \cite{Chen:1998awa} and anomaly-mediated \cite{Randall:1998uk,Thomas:1998wy} SUSY breaking, natural weak-scale RPV SUSY \cite{Hewett:2004nw,Barbier:2004ez,Graham:2012th}, twin Higgs \cite{Chacko:2005pe}, 
hidden valley models \cite{Han:2007ae}, and others such as \cite{Falkowski:2014sma}. There has also been the consideration of cosmological motivations for displaced vertices arising from the requirement that low-scale baryon number violating interactions are sufficiently weak  that they do not wash out a pre-existing baryon asymmetry generated at higher scales \cite{Barry:2013nva}.  

\begin{figure*}[t]
\begin{center}
\subfigure[]{\includegraphics[width=0.495\textwidth]{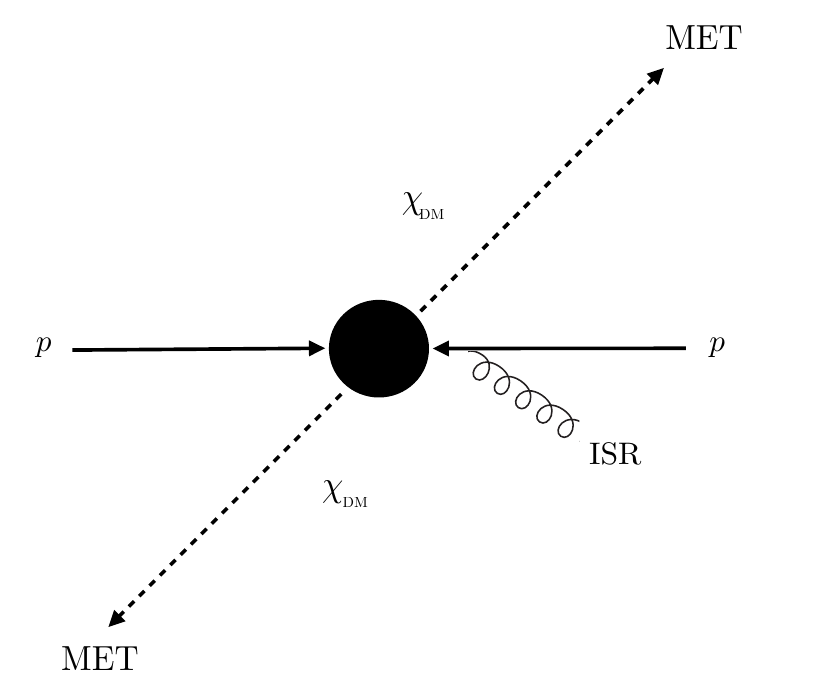}}
\subfigure[]
{\includegraphics[width=0.45\textwidth]{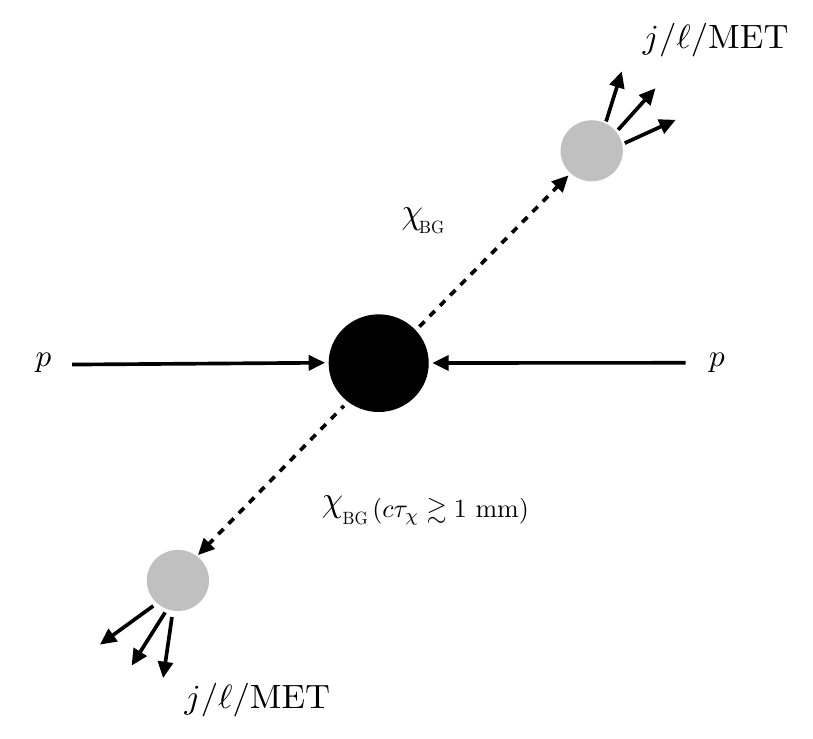}}
\caption{Schematic diagram showing the pair-production at the LHC of (a) dark matter in stable WIMP dark matter searches, with associated initial state radiation (ISR); (b) the analogous production of the meta-stable WIMP triggering baryogenesis, which decays at a displaced vertex to jets, leptons, and/or missing transverse energy.}
\label{fig:WIMP_schematic}
\end{center}
\end{figure*}

It is reasonable to expect that fields involved in the low-scale baryogenesis scenarios described above can be produced at the LHC. The baryon parent, $\chi$, must have a very small decay width, which can be the result of an approximate $Z_2$ (or larger) symmetry. The decay of $\chi$ is suppressed by the symmetry-breaking parameter, while its production may proceed through symmetry-preserving processes with larger rates. This is true in familiar examples with discrete symmetries, such as the lightest supersymmetric (SUSY) particle (LSP) in SUSY theories with approximate R-parity. This leads to important conclusions:~the production rate of $\chi$ at the LHC or other colliders can be more significant than na\"ively expected from its long lifetime (but still potentially small), and $\chi$ is typically produced in pairs, leading to signatures with two displaced vertices. Therefore, in situations where large backgrounds are possible (such as all-hadronic DVs), requiring two DVs per event can dramatically suppress backgrounds with only a small loss in signal efficiency, enhancing the sensitivity to signals with small cross sections. We illustrate such cosmologically motivated collider searches in Fig.~\ref{fig:WIMP_schematic}.

In this work, we focus on a concrete, viable mechanism that realizes the above scenario: the recently proposed, novel low-scale baryogenesis mechanism in which the baryon asymmetry is generated by the out-of-equilibrium decay of a meta-stable WIMP after its thermal freeze-out \cite{Cui:2012jh}. For convenience, we refer to this mechanism as ``WIMP baryogenesis''. In WIMP baryogenesis, a meta-stable particle $\chi$ undergoes thermal freeze-out, rendering a relic abundance in accordance with the conventional WIMP miracle. After freeze-out, $\chi$ decays via baryon or lepton-number-violating interactions; in the presence of CP violation, this leads to the creation of a baryon asymmetry. The key stages of WIMP baryogenesis are summarized in Fig.~\ref{cartoon}. Since $\chi$ decays far out of equilibrium, WIMP baryogenesis automatically implies $c\tau_\chi\gtrsim$ mm, while the reverse of the annihilation processes responsible for freeze-out can give a portal for producing the long-lived WIMPs at colliders.  

In addition to its role as a concrete implementation of a weak-scale baryogenesis model giving rise to displaced vertices, WIMP baryogenesis is a unique baryogenesis mechanism that naturally gives a robust prediction for the baryon abundance around the observed value, based on a generalized ``WIMP miracle''.  As a low-scale mechanism, it offers a viable path for baryogenesis in scenarios where a high-scale baryon asymmetry would be diluted \cite{Acharya:2009zt} or washed out \cite{Barry:2013nva}. We can compactly estimate the present-day baryon  abundance $\Omega_{\Delta{B}}$ with a just few parameters: $\Omega_{\Delta{B}}=\epsilon_{CP}\Omega_\chi^{\infty}$, where $\epsilon_{CP}$ is the baryon asymmetry produced per decay, and $\Omega_\chi^{\infty}$ would be the $\chi$ relic abundance if it were a stable WIMP. The baryon asymmetry therefore has a WIMP-miracle-like abundance. Assuming DM is a different WIMP that is stable WIMP, this mechanism can naturally address the similarity between the present-day DM and baryon abundances based on a shared WIMP miracle, while intrinsically including a mechanism for generating a baryon asymmetry. There have been several concrete implementations of WIMP baryogenesis, including in minimal, mini-split SUSY models \cite{Cui:2013bta}, where the bino is the meta-stable WIMP responsible for baryogenesis, as well as extended natural SUSY models \cite{Cui:2012jh} and other examples \cite{RompineveSorbello:2013xwa, Arcadi:2013jza}.

A review of the DV search status at LEP, the Tevatron, and earlier LHC runs can be found in \cite{Graham:2012th}. Both ATLAS and CMS  have excellent tracker resolution and have recently made impressive progress on improving DV search sensitivities in various channels\footnote{For instance, significant improvement in limits and sensitivities have been achieved in the past two years since the publication of Ref.~\cite{Graham:2012th}.}. The exclusion limits placed in the DV search channels by the LHC analysis have surpassed any previous searches for particle masses $\gtrsim$100 GeV for pair production, and these are expected to improve in future runs. The LEP2 searches may have competitive sensitivity to particles within its kinematically accessible range (below 100 GeV each for pair-produced particles), but are limited by the total luminosity, $0.6~\rm fb^{-1}$. Furthermore, one of the most sensitive searches, which is by ALEPH \cite{Barate:1998zp}, requires events with only a few (2-6) tracks, excluding the generic case of displaced jets with many tracks. Therefore, for the main results of our paper, we focus on two particular searches relevant for WIMP baryogenesis: the CMS search for displaced dijets \cite{CMS-PAS-EXO-12-038}, and the ATLAS search for a displaced muon+tracks \cite{ATLAS-CONF-2013-092}. Both of these have very low backgrounds and already impose limits in the $O(0.1)$ fb range based on 8 TeV $\sim20~\rm fb^{-1}$ data, and these can serve as sensitive probes of WIMP baryogenesis.
We estimate current limits on DV signals for models of WIMP baryogenesis from the recent LHC searches and their possible extensions at 13 TeV, highlighting gaps and proposing improved search strategies. In particular, we identify  two representative simplified models that exemplify the phenomenology of WIMP baryogenesis models:~in the first, we consider new fields charged under SM gauge interactions, which typically feature large production cross sections and sensitivity to masses of a few TeV, while in the second, we consider SM gauge-singlet fields which generally have small production cross sections even for 100 GeV masses. We show that, for the experimental analyses considered in this paper, strategies such as tagging two DVs with the possibility of relaxing other requirements can yield improvements in sensitivity to signal cross sections by up to two orders of magnitude.

The outline for the rest of the paper is as follows: in Section \ref{sec: simplified_models}, we briefly review the mechanism of WIMP baryogenesis, and we develop a simplified model approach based on effective operators to identify two benchmark models that represent a large class of models. In Section \ref{WIMPbg_LHC_general}, we discuss general aspects of DV searches at the LHC, provide more details of the ATLAS and CMS DV analyses, and describe our methods of Monte Carlo (MC) simulation. In Section \ref{8TeV_limits}, we estimate the constraints on our benchmark models from the ATLAS DV muon+tracks and CMS DV dijet searches based on 8 TeV, 20 fb$^{-1}$ data, and in Section \ref{13TeV_prospect} we discuss the projected limits at future 13 TeV LHC runs, proposing strategies to improve sensitivity, such as tagging two DVs\footnote{Such 2 DV analyses should complement the existing single DV analyses, which are necessary to maintain sensitivity to final states that are sufficiently long-lived that only one WIMP decays inside the detector volume.} and lowering kinematic thresholds where possible to enhance sensitivity to low-mass, long-lived particles. Finally, Section \ref{concl} concludes our discussions.

\section{Models of WIMP Baryogenesis at Colliders}\label{sec: simplified_models}
\subsection{Review of WIMP baryogenesis theory}\label{subsec: WIMPBG}

\begin{figure}
   \begin{center}
        \includegraphics[height=80mm]{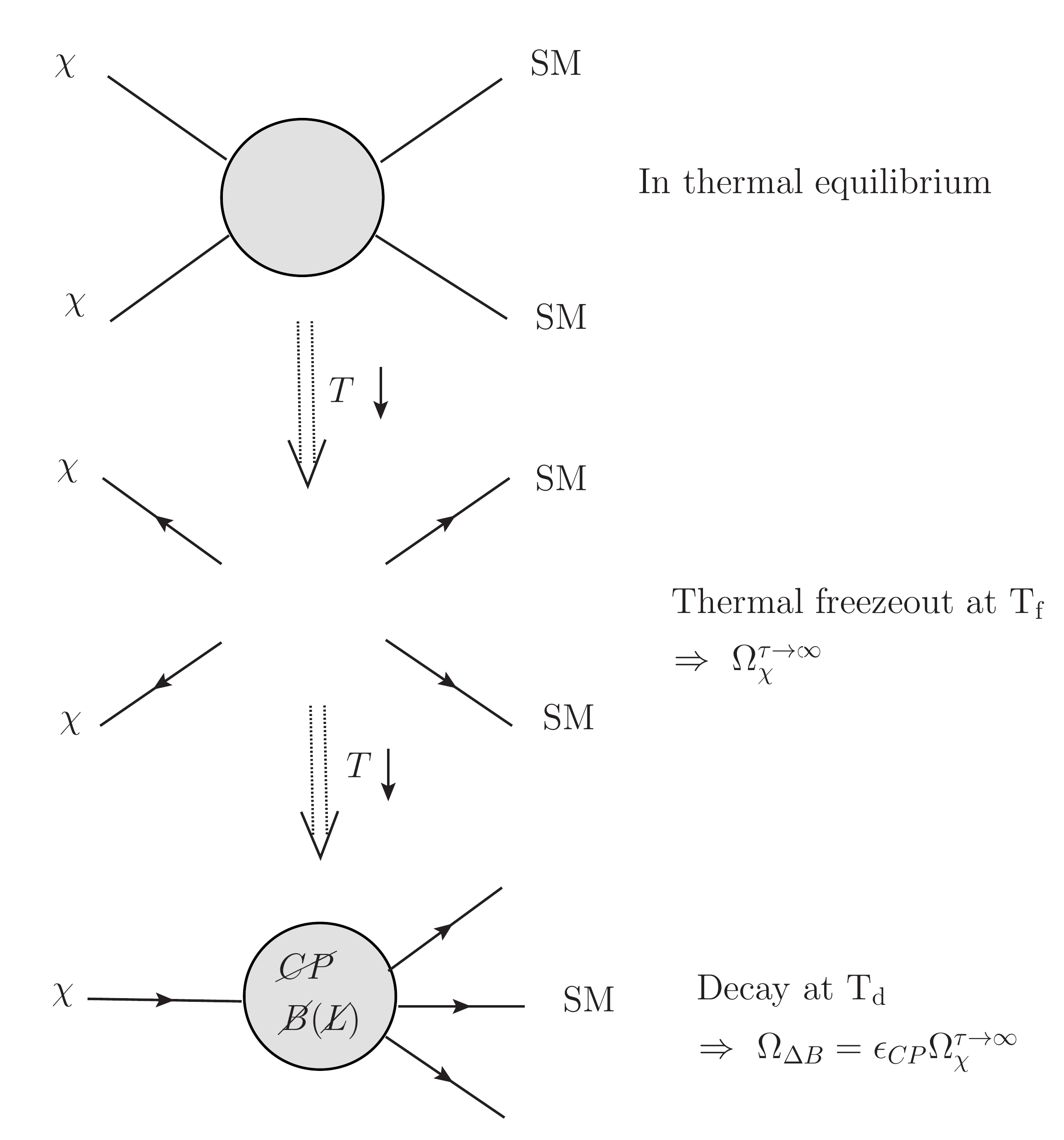} 
     \end{center} 
        \caption{Illustration of the key processes in WIMP baryogenesis via the out-of-equilibrium decay of a meta-stable WIMP $\chi$. The dashed double arrow indicates the arrow of time, along which the temperature decreases.}  
        \label{cartoon}   
\end{figure}

WIMP baryogenesis is an example of a weak-scale baryogenesis mechanism that predicts displaced vertices at colliders. In WIMP baryogenesis, the baryon asymmetry is generated by the out-of-equilibrium decay of a meta-stable WIMP after its thermal freeze-out \cite{Cui:2012jh}. This occurs if the WIMP is sufficiently long-lived and decays through interactions that violate baryon or lepton number, as well as the C and CP symmetries. The processes leading to WIMP baryogenesis and an estimate for the baryon asymmetry are shown in Fig.~\ref{cartoon}. The original paper \cite{Cui:2012jh} demonstrated an embedding of the general concept of WIMP baryogenesis in a model of natural SUSY with R-parity violation (RPV), naturally resolving the issue that RPV interactions may erase $\Omega_{\Delta B}$ generated by high-scale baryogenesis \cite{Barbier:2004ez, Barry:2013nva}. Another realization of WIMP baryogenesis was proposed in the Minimal Supersymmetric Standard Model (MSSM) with a mini-split spectrum \cite{Cui:2013bta}. Here, the bino was shown to be a suitable candidate for the meta-stable WIMP responsible for baryogenesis. Other  examples of WIMP baryogenesis have since been proposed \cite{RompineveSorbello:2013xwa, Arcadi:2013jza}.

For our purposes, the most intriguing feature of  WIMP baryogenesis models is that the long proper lifetime of the WIMP baryon parent 
implies that its decay   either generates a displaced vertex within the detector when produced at a collider, or it escapes the detector entirely, in which case the final state could be detected as missing energy. To see this, recall that the meta-stable WIMP $\chi$ needs to live beyond its thermal freeze-out time, which is around the weak scale. This is equivalent to requiring the proper lifetime to satisfy
\be
\tau_\chi\gtrsim t_{\rm fo}=0.3g_*^{-1/2}\frac{M_{\rm pl}}{T_{\rm fo}^2}\sim \left(\frac{T_{\rm fo}}{\rm 100 \,\,GeV}\right)^{-2}10^{-10}\rm \,\,sec,
\ee
where the subscript ``fo'' denotes freeze-out. For thermal WIMP DM $\chi$, obtaining the correct relic abundance requires $M_\chi/T_{\rm fo}\sim20$; in WIMP baryogenesis, 
the freeze-out temperature can be somewhat higher, due to the fact that a small CP asymmetry factor $\epsilon_{\rm CP}$ and mass ratio $m_p/m_\chi$ can necessitate an over-abundance of the meta-stable WIMP abundance after freeze-out; nevertheless, $T_{\rm fo}$ is always around the weak scale. Note that even in the case where the proper decay length is larger than the detector scale and predominantly gives a missing energy signature, there can still be a fraction of $\chi$ decaying inside the detector. Since most DV signatures have no irreducible SM backgrounds, unlike missing energy searches, the DV decay modes can still be a useful search channel, even when it is not the leading mode in terms of signal rate.

We have argued that the baron parent, $\chi$, has a long lifetime. There are, however, typically other particles in the spectrum at a comparable mass to $\chi$ which can also be long-lived. When a baryon asymmetry results from the decay of an out-of-equilibrium particle, the requisite CP asymmetry arises from the interference of tree and loop diagrams; in particular, the physical CP phase and additional B-violating source beyond tree-level  required for baryogenesis by the Nanopoulos-Weinberg theorem \cite{Nanopoulos:1979gx}  necessitates other states that can appear on-shell in the loop. These states are typically also charged under the approximate $Z_2$ symmetry with a long lifetime, and can also be produced at colliders. As an example, the mini-split SUSY model of \cite{Cui:2013bta} predicts a long-lived wino or gluino in the spectrum that is lighter than the baryon parent, which is the bino, in order to generate a CP asymmetry. Therefore, the out-of-equilibrium Sakharov condition predicts a variety of possible long-lived states giving rise to displaced vertices. As further speculation, an exciting yet very challenging next step would be to measure the CP-violating effect responsible for baryogenesis directly from the charge asymmetry in the final-state system. This would demand very high luminosity and dedicated search strategies, and is beyond the scope of this work.

\subsection{Simplified models for collider studies}\label{subsec: simp_model}

The models of WIMP baryogenesis proposed to-date \cite{Cui:2012jh,Cui:2013bta,RompineveSorbello:2013xwa, Arcadi:2013jza} share some common features, such as the long lifetime of the particle(s) responsible for baryogenesis leading to displaced vertices and/or missing energy signatures, although they differ in model details. While there are many potentially interesting avenues to pursue in building viable models of WIMP baryogenesis, the resulting LHC phenomenology is insensitive to all but a few parameters: the production mechanisms or gauge charges of the meta-stable WIMPs, their decay channels, and the proper lifetime set by the approximate breaking of the $Z_2$ symmetry. Other details of the baryogenesis model, such as the precise CP asymmetry, appear as higher-order corrections to WIMP baryogenesis processes at the LHC and are challenging to probe. Therefore, phenomenological studies of baryogenesis at the LHC are well-suited to the simplified model  approach \cite{Alves:2011wf} which is now in common use for supersymmetric and dark matter constraints at the LHC. We employ this simplified models approach, assuming that any given WIMP baryogenesis model is most easily probed via a single long-lived particle $\chi$, and we classify its phenomenology according to its production, decay, and lifetime.

We focus on new Majorana fermions as the long-lived states responsible for baryogenesis, which allows a simple realization of CP- and $B/L$-number violation. There are also typically other light Majorana fermions in the spectrum as described in Section \ref{subsec: WIMPBG}. For phenomenology studies here, we focus on this set of new long-lived states predicted by WIMP baryogenesis, and single out the most interesting field, namely the state $\chi$ with the largest LHC production cross section. All other particles in the spectrum are assumed to be decoupled;  their presence is only a minor perturbation on the simplified model.\\

\noindent {\bf Production:} The $\chi$ production mechanism must respect the $Z_2$ (or larger) symmetry that makes it meta-stable, which implies that $\chi$ should be produced in pairs at the LHC. We expect at least one relatively large coupling to allow $\chi$ to thermally freeze out with a (near) DM abundance, which can give a sizeable production rate at colliders by reversing the annihilation diagram. As a Majorana fermion, $\chi$ is either a SM singlet or in a real representation of the SM gauge group. The two simplest possibilities we consider in our paper are:

\benum

 \item \textbf{New meta-stable states in adjoint representation of SM gauge group}\\
 $\chi$ can be directly pair-produced via $s$-channel exchange of a SM gauge boson, and its production cross section is fixed by its mass and SM gauge coupling. The production part of its Lagrangian is
 \be
 \mathcal{L}_{\rm prod}^{\rm adjoint} \supset i\chi^\dagger \bar\sigma^\mu D_\mu \chi,
 \ee
 where $D_\mu$ is the relevant gauge covariant derivative. This is similar to the production of gluinos and winos in SUSY models, and a WIMP baryogenesis model with such light Majorana adjoint fermions is realized in the mini-split SUSY model with decoupled sfermions and Higgsinos \cite{Cui:2013bta}. Due to the sizeable $\chi$ pair-production cross section through gauge interactions, we expect that the LHC has sensitivity to large $\chi$ masses in this scenario.
   
 \item \textbf{Meta-stable new states that are singlets} \\
The production of gauge-singlet states is highly model-dependent. We  proceed by analogy with DM searches, in which it is common to classify $\chi$ production by its effective interactions with the SM. The lowest-dimension operators are of the form:
\be\label{eq:Lprod_singlet}
\mathcal{L}_{\rm prod}^{\rm singlet} \supset \frac{c_H}{\Lambda_H}\chi^2 |H|^2 + \frac{c_q}{{\Lambda_q^2}}(\bar\chi\Gamma\chi)(\bar q\Gamma' q) + \frac{c_g}{{\Lambda_g^3}}\chi^2(G_{\mu\nu})^2+\ldots,
\ee
where the first term is the Higgs portal, the second term shows a schematic coupling to quarks via Dirac matrix structures $\Gamma, \Gamma'$, and the third term is the gluon portal. Singlets can also be produced from decays of heavier particles; in the spirit of the simplified model, however, we only consider direct production, remarking that $\chi$ production via cascade decays will have stronger bounds than  we present. For concreteness we focus on the lowest-dimensional, Higgs-portal coupling. Other operators lead to similar collider phenomenology, but with different production rates.

\eenum

\noindent{\bf Decay:} In contrast with the production of the meta-stable state $\chi$, which is dominated by $Z_2$-preserving interactions, the decay of $\chi$ is mediated by $Z_2$-breaking interactions.  The decay modes of $\chi$ should also violate baryon and/or lepton number because it is responsible for baryogenesis. The lowest-dimensional interactions of this type allow $\chi$ to decay into three SM fermions. We classify all SM fermion trilinear operators that can couple to $\chi$ in either the adjoint or singlet representation of the SM gauge groups. Our notation is as follows:~operators labelled by ``$q$'' contain at least two baryons/quarks;  operators labelled by ``$\ell$'' contain only leptons; operators labelled by  ``$\rm L$'' signals that the operator contains two left-handed doublet fields; ``$\rm R$'' operators contain only singlets under $\mathrm{SU}(2)_{\rm L}$. The effective Lagrangian for $\chi$ decay is:
\be\label{eq:operator_coupling}
\mathcal{L}_{\rm decay} \supset \chi(\mathcal{O}, \,\,\mathcal{O}^\dagger) + \rm h.c,,
\ee
where the possible operators $\mathcal{O}$ are
\bi
\item $\mathcal{O}_{q\mathrm{L}}$: $\lambda_{ijk}'~Q_iD_jL_k$, $\eta_{ijk}'~Q_iU_jL_k^\dag$, $\eta_{ijk}''~Q_iQ_jd_k^{\dag}$,
\item $\mathcal{O}_{q\mathrm{R}}$: $\lambda_{ijk}''~U_iD_jD_k$, $\eta_{ijk}~U_iE_jD_k^\dag$,
\item $\mathcal{O}_{\ell\mathrm{L}}$: $\lambda_{ijk}~L_iL_jE_k$,
\ei 
depending on the gauge charges of $\chi$. If $\chi$ is a singlet, it can couple to all of these operators. If instead $\chi$ is in the adjoint representation of the SM gauge field, it can only couple to operators containing fields transforming in the fundamental representation of the corresponding gauge interaction. For example, an $\mathrm{SU}(3)$ octet (like a gluino) could couple to $\mathcal{O}_{q\mathrm{L}}$ and $\mathcal{O}_{q\mathrm{R}}$, while an $\mathrm{SU}(2)_{\rm L}$ triplet (like a wino) could couple to $\mathcal{O}_{q\mathrm{L}}$ and $\mathcal{O}_{\ell\mathrm{L}}$. Many of these trilinear operators are familiar from theories of RPV SUSY, where we have included  terms corresponding to both holomorphic superpotential and non-holomorphic K\"ahler potential interactions \cite{Csaki:2013jza}; of course, in a non-SUSY theory, there is no distinction between these sets of operators. Note that the operators have non-trivial flavour structures. In this work we focus on the simplest case where the final states are light quarks/leptons (first two generations). Final states involving third generation fields such as top quarks lead to more complex topologies and are in general more strongly constrained.  

We make a few comments on the effective operator approach. First, this approach obscures the possibility that ultraviolet (UV) completions of the above operators can introduce additional light particles, allowing $\chi$ to decay to on-shell intermediate states, e.g. $\chi\rightarrow \tilde{\bar{t}}{t}\rightarrow jj ~t$ as in \cite{Cui:2012jh}. In general, having on-shell intermediate states makes the theory more easily discovered through resonance tagging, but otherwise does not substantially alter the phenomenology. Second, we have listed both baryon and lepton number violating operators. However, if $\chi$ decays after the electroweak phase transition  in the early universe, only baryon violating operators can give rise to successful baryogenesis due to the inefficacy of $(B+L)$-violating sphaleron interactions. Therefore, baryon violating operators are particularly well-motivated for $\chi\sim O(100)$ GeV, giving all hadronic final states for collider searches, while $\chi$ well above the weak scale can decay into either a baryon or lepton asymmetry.

 \subsection{Two case studies}
 \label{sec:benchmarks}

Even in the simplified model approach, there is still a vast multiplicity of possible $\chi$ production and decay modes. To focus our studies, we consider two representative benchmark simplified models for $\chi$ production. Our benchmark models each cover one of the production classes in Section \ref{subsec: simp_model}:~in one model, $\chi$ is charged under SM gauge interactions, while in the other, $\chi$ is a singlet produced through the Higgs portal. With these models, we gain an understanding of the range of masses and cross sections to which the LHC is sensitive.\\

\noindent \textbf{MSSM wino with RPV couplings}\\
The simplest example of a Majorana fermion charged in a real representation of the SM gauge group is an adjoint fermion. This is directly analogous to the charges of the gaugino fields in the MSSM. After integrating out the decoupled sfermion fields, the MSSM with RPV also generates the baryon and lepton number violating effective operators listed below Eq.~(\ref{eq:operator_coupling}). In particular, we take our $\chi$ to be a wino:~the bounds on a gluino $\chi$ are only going to be stronger than the wino due to the large production rate, and the DV searches are slightly complicated by the gluino R-hadron state that yields tracks leading from the primary vertex to the displaced vertex.

Because the wino production cross section is fixed by the SM electroweak coupling, the LHC sensitivity to a wino is parameterized by its mass. As we will show, the bounds on wino pair production with RPV displaced decays  at 8 TeV are already in the $600-1000$ GeV range, and the wino mass reach can be up to 2.5 TeV with decay lengths $\sim$ cm and $O(1000)~ {\rm fb}^{-1}$ luminosity. This is in contrast with a reach of $\lesssim$ TeV for a wino which decays promptly to gauge bosons + MET (for example, see \cite{Baer:2012ts}), or $\lesssim$ few hundred GeV for the most minimal models that give disappearing charged track signatures \cite{Ibe:2006de,Buckley:2009kv}.

We consider the pair production of wino-type charginos and associated chargino-neutralino production, i.e., $q\bar{q}\rightarrow \chi^+\chi^-$ through $Z^0$ and $q\bar{q}'\rightarrow\chi^0\chi^{\pm}$ though $W^{\pm}$. With winos, there is a minor complication in that the charginos can decay into neutralinos, while both can decay through the RPV operators into quarks and leptons. In our analyses, we make the simplifying assumption that $c\tau_{\tilde\chi^\pm\rightarrow\tilde\chi^0}\ll c\tau_{\tilde{\chi}^\pm\,\mathrm{(RPV)}}$, so that  tracks at the displaced vertex come entirely from neutralinos and the  track coming from the soft pion emitted by chargino decay is not associated with any vertex. In reality, it may be that $c\tau_{\tilde\chi^\pm\rightarrow\tilde\chi^0}\sim c\tau_{\tilde{\chi}^\pm\,\mathrm{(RPV)}}$, in which case the DVs result from a mixture of charginos and neutralinos; this gives an added complication of a charged track leading from the primary vertex to the secondary vertex, but this track may not be properly reconstructed as it disappears in the middle of the detector. The presence of this additional track  does not appear to substantially affect the vertex selection efficiency for either of the ATLAS or CMS analyses we consider, and we ignore this complication for simplicity.    \\

 \noindent \textbf{Singlet coupled via Higgs Portal}
 
 \noindent The lowest-dimension operator coupling a singlet $\chi$ to the SM is the Higgs portal, which is the first term in Eq.~(\ref{eq:Lprod_singlet}). A simple and natural UV completion of this interaction is with a singlet scalar $S$ that mixes with the SM Higgs. If $S$ has a Yukawa coupling to a pair of $\chi$, this leads to the Higgs-portal production of $\chi$ via the exchange of a single SM-like Higgs boson after mixing ($gg\rightarrow h^*\rightarrow\chi\chi$). $\chi$ can also be produced resonantly through the $S$-like state, and this may be a more important production mechanism if $S$ is within reach of the LHC or another collider; this possibility was mentioned in \cite{Cui:2012jh}. Since the LHC has established the existence of a 125 GeV SM-like Higgs, while the other possibilities are more model-dependent, we assume the minimal spectrum where the $S$-like scalar is heavy and decouples, and focus on the production channel via the SM Higgs portal.  
 
The coupling between the SM-like Higgs state $h$ and the meta-stable WIMP singlets $\chi$ after taking into account the mixing is:
 \beq
 \mathcal{L}\supset \frac{\lambda_{S\chi\chi}}{2}\sin\alpha~ h\chi\chi,
 \eeq
 where $\alpha$ is the $H-S$ mixing angle and $\lambda_{S\chi\chi}$ is the original $S\chi\chi$ coupling.  This model can be parametrized by two quantities:~the effective coupling $\lambda_{S\chi\chi}\sin\alpha$, and the WIMP mass $m_\chi$. Because the production of $\chi$ through the Higgs portal has a factor of $\cos\alpha$ from the SM-like $ggh$ coupling and $\sin\alpha$ from the $h\chi\chi$ coupling, the rate actually depends on $\sin2\alpha$. The current model-independent limits on the singlet fraction of the SM-like Higgs observed at 126 GeV are currently quite weak, with $\lambda_{S\chi\chi}\sin2\alpha\lesssim1$ \cite{Farina:2013fsa}, although the constraints can be stronger depending on the model.
 
For $\chi$ production through the Higgs portal, we focus on the regime $m_\chi > m_h/2$. In the region $m_\chi<m_h/2$,  the dominant production mechanism is through an on-shell Higgs, and the production of $\chi$ can be copious at the 8 or 13 TeV LHC ($\sigma_{gg\rightarrow h\rightarrow \chi\chi}\sim (\lambda_{S\chi\chi}\sin2\alpha)^2\cdot O(10)$ pb), and the constraints are correspondingly strong, even after accounting for the lower experimental acceptance associated with the softer objects coming from the decay of light $\chi$. There are also indirect constraints on the non-SM decay branching fraction of Higgs based on global fits of data \cite{Englert:2011yb, Englert:2011aa}. On the other hand, the Higgs is off-shell if $m_\chi > m_h/2$ and the signal rate falls rapidly with increasing $m_\chi$, even for large mixing. Displaced vertex searches, which typically feature low backgrounds, offer the best hope of probing such new physics through the off-shell Higgs portal, even though they are challenging to conduct; in Section \ref{13TeV_prospect}, we propose ways of enhancing this sensitivity. By contrast, it has been shown that when the Higgs is off-shell (even when close to resonance), the signal is inaccessible at the LHC if $\chi$ decays invisibly \cite{Kanemura:2010sh}, although future lepton colliders such as TLEP or the ILC may moderately improve the reach up to $\sim100$ GeV masses with large mixing in the invisible decay channel \cite{Matsumoto:2010bh, Kanemura:2011nm,Chacko:2013lna}. Similarly, most prompt decay scenarios are  overwhelmed by SM backgrounds and difficult to probe in the off-shell Higgs regime.

In principle, there are three independent parameters in the Higgs portal model:~$m_\chi$, $c\tau_\chi$, and $\lambda_{S\chi\chi}\sin2\alpha$. As we  show in Section \ref{13TeV_prospect}, it will require very high luminosity for the LHC to probe $m_\chi\approx100-150$ GeV, and the prospects are poor for even heavier $\chi$, so we restrict ourselves to a single benchmark point $m_\chi=150$ GeV, scanning over different values of $\lambda_{S\chi\chi}\sin2\alpha$. This mass point is also convenient to study since it satisfies $m_\chi<m_t$, and so the Higgs Effective Field Theory (HEFT) model in Madgraph 5 allows Monte Carlo event generation without being strongly affected by top-mass effects. We have confirmed that the $\chi$ production rate is inaccessibly small for much heavier $m_\chi$ by analytically computing the leading-order (LO) cross section at 13 TeV based on the full one-loop matrix element of the the $ggh$ vertex given in \cite{Spira:1995rr,ggh_LO}, combined with the proper phase space factors and convolved with the MSTW2008 PDF \cite{Martin:2009iq}. We find that, for $\lambda_{S\chi\chi}\sin2\alpha=1$, the cross section falls from $\approx 1$ fb at $m_\chi=150$ GeV, to $0.05$ fb at $m_\chi=300$ GeV. Thus, well above 150 GeV, the cross section is so small that very few events remain after selection, even for unreasonably large  couplings.

 \section{Studies of ``Displaced'' WIMP Baryogenesis at the LHC}\label{WIMPbg_LHC_general}

We can summarize the generic features of  expected collider signatures from baryogenesis via the decay of meta-stable weak-scale particles:
\bi
\item There exist new, long-lived weak-scale particles with proper decay length $\gtrsim O(1)$ mm, giving rise to displaced vertices.
\item The long-lived nature of such particles is typically due to some approximate $Z_2$ (or larger) symmetry, such as R-parity in SUSY. Therefore, the long-lived particles are produced in pairs, and we expect two displaced vertices in a typical signal event.
 \item Depending on the decay channel, the final states originating from the displaced vertices include various baryon and lepton number violating topologies:~three jets, dijets+charged lepton, dijets+MET, dileptons+MET, and more complicated states when the long-lived particle decays to a top quark or a $\tau$. 
  \ei

In this section, we first describe the relevant DV searches at the LHC for WIMP baryogenesis. We focus on searches where the long-lived particle is reconstructed inside of the tracking system:~these are sensitive to several orders of magnitude of particle lifetimes at the lower limit predicted by weak-scale baryogenesis. These searches are also amenable to Monte Carlo simulation. Focusing on two characteristics searches, one for all-hadronic final states in CMS and the other with leptons in ATLAS, we describe the aspects of the searches that are crucial for studies of WIMP baryogenesis and how these pertain to our models. 

In Section \ref{sec:MC}, we discuss the Monte Carlo (MC) methods we developed to apply these searches to the phenomenology of the two benchmark simplified models of WIMP baryogenesis from Section~\ref{sec:benchmarks}. For the reader interested in skipping to current and projected limits, these can be found in Sections \ref{8TeV_limits} and \ref{13TeV_prospect}.

 \subsection{Existing LHC DV searches}
The LHC experiments have released several  searches for long-lived particles (among the most relevant for us Refs.~\cite{CMS-PAS-EXO-12-038,ATLAS-CONF-2013-092,ATLAS-CONF-2013-069,Chatrchyan:2012jna,Khachatryan:2014mea,ATLAS:2012av,ATLAS-CONF-2014-041}), with the intention of continuing this work at 13 TeV. These are classified according to the particles in the final state, and the part of the detector in which the long-lived particle decays, with different analyses covering decays ranging from the inner detector to the outer muon system. We focus on analyses where the decays occur in the inner tracking system, both because this is the expected range for WIMP baryogenesis or other out-of-equilibrium weak-scale baryogenesis scenarios, and because they cover a wide range of displacements $c\tau\sim O(0.1-100)$ cm, with both experiments having relatively good vertex reconstruction capabilities over this range.  Since displaced vertex reconstruction is quite good in both experiments, these searches are also well-modelled by MC simulation. The best cross section limits for searches in the tracker are currently in the fb range. There also exist searches for longer-lived WIMPs of mass $\sim100$ GeV that decay inside the hadronic calorimeter (HCAL) \cite{ATLAS-CONF-2014-041} or muon system \cite{ATLAS:2012av}; the larger background and/or poorer reconstruction of events typically lead to cross section limits in the pb range.

 In terms of final states, we consider two existing LHC searches that are sensitive to  many of the final state topologies expected from Section \ref{sec: simplified_models}:~a CMS search for displaced dijets, and an ATLAS search for a displaced muon and tracks. There is a search for displaced leptons from CMS probing purely leptonic final states, which for instance can result from a $\chi LLE^{\rm c}$ operator; however, they do not provide the same efficiency information needed for our MC methods described in Section \ref{sec:MC}. Therefore, we defer a recasting of this study to future work, although we comment that since both the displaced dijet and muon+tracks analyses constrain a similar range of masses and production cross sections for WIMP baryogenesis, we expect the dilepton analysis will also give comparable constraints. 

In this section, we detail the important aspects of each search, highlighting the observables and methods used that are important for constraining models of WIMP baryogenesis.

\subsubsection{CMS displaced dijet analysis}
\label{sec:CMS_intro}
The CMS analysis \cite{CMS-PAS-EXO-12-038} looks for long-lived particles decaying into dijets, in particular focusing on the example of $gg\rightarrow H^0\rightarrow SS,\, S\rightarrow q\bar{q}$, where $H^0$ is a heavy Higgs-like scalar, and $S$ is a long-lived scalar decaying to dijets. The dominant backgrounds come from quantum chromodynamics (QCD) multijet processes where there are random crossings/mis-reconstruction of tracks and where there is an interaction of hadrons from the primary vertex with material in the detector. Backgrounds from $b-$ and $c-$hadrons are suppressed by requiring a transverse impact parameter of $>0.5$ mm for displaced tracks (which is sufficient to exclude most $b$-hadron tracks), vetoing jets with more than one prompt track, and vetoing jets where more than $\sim10\%$ of the jet energy is carried by prompt tracks. Furthermore, it is rare for displaced vertices from heavy-flavour processes to produce two hard jets from the single displaced vertex. The cut on impact parameter limits the efficacy of the search for particles with characteristic transverse decay distance of $\lesssim0.5$ mm.

Because the final state in the CMS search is entirely hadronic, it relies on a dedicated trigger for displaced jets, where a ``displaced jet'' is a jet with most of its energy carried by particles with a large transverse impact parameter relative to the primary vertex. The trigger requires $H_{\rm T}>300$ GeV and at least two displaced jets with $p_{\rm T}>60$ GeV, and  the sensitivity of the analysis is significantly reduced for light, long-lived particles that produce softer jets.

\begin{figure*}[t]
\begin{center}
\subfigure[DV track clustering for $S\rightarrow2j$]{\includegraphics[width=0.3\textwidth]{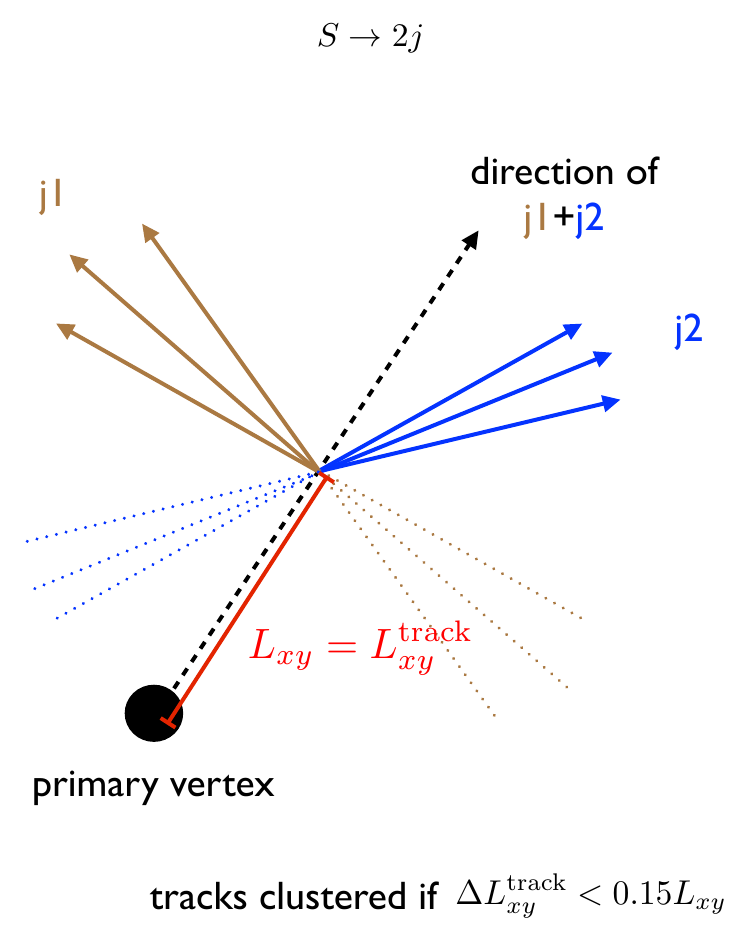}}\hspace{3cm}
\subfigure[DV track clustering for $\chi\rightarrow3j$]{\includegraphics[width=0.33\textwidth]{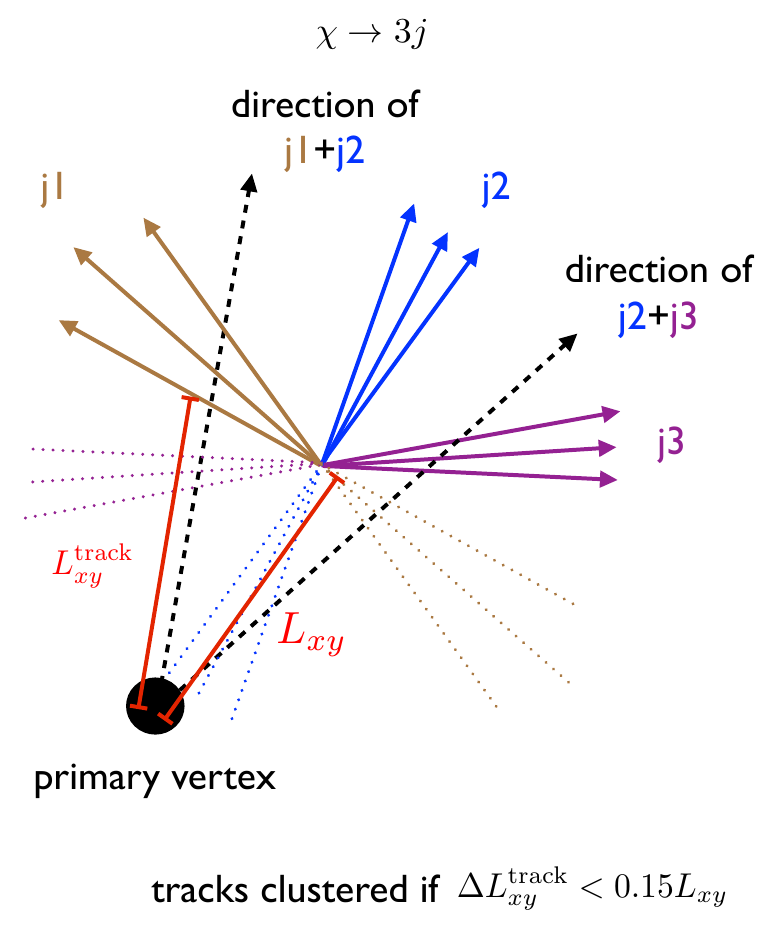}}
\caption{Schematic diagram showing the procedure for clustering tracks in the CMS displaced dijet analysis. The point of intersection between the dijet momentum and each track is determined, with the distance from the primary vertex defined as $L_{xy}^{\rm track}$. Tracks are clustered if the distance between $L_{xy}^{\rm track}$ is less than 15\% of $L_{xy}$, the distance from primary vertex to DV. }
\label{fig:cluster_schematic}
\end{center}
\end{figure*}

The analysis reconstructs displaced vertices by finding the large-impact-parameter tracks associated with the displaced jets and attempting to fit a common vertex to them with transverse distance $L_{xy}$. A variety of track and vertex selection criteria are applied. Vertices are rejected if more than one associated track is prompt and if the prompt energy fraction is above $\approx10\%$. Finally, it is assumed that since the long-lived particle decays into dijets, the dijet momentum should reconstruct the momentum of the long-lived particle. A line is drawn from the primary vertex in the direction of the dijet momentum, and each track intersects this line at a particular point, which has transverse distance to the primary vertex $L^{\rm track}_{xy}$; the tracks are ``clustered'' together if their intersection points are closer than 15\% of the distance from the primary vertex to the secondary vertex (see Fig.~\ref{fig:cluster_schematic}). When the dijet momentum coincides with the long-lived particle momentum, as in the case of a dijet decay, then the tracks should all intersect the dijet momentum line at the position of the displaced vertex, and should hence be clustered together. 

\begin{figure*}[t]
\begin{center}
\subfigure[Impact parameter projection for $S\rightarrow2j$]{\includegraphics[width=0.35\textwidth]{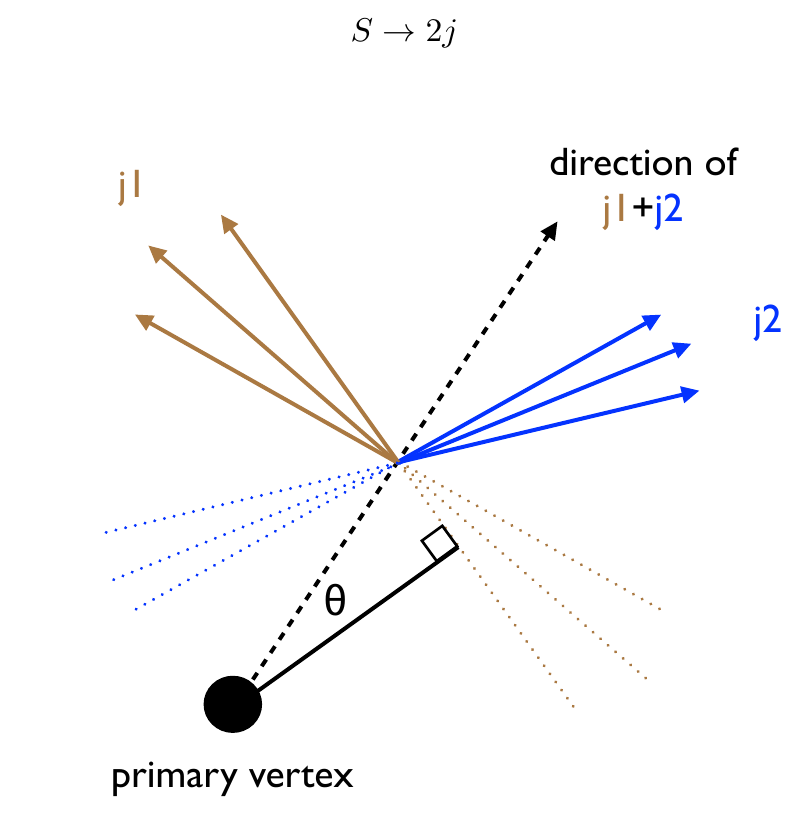}}\hspace{3cm}
\subfigure[Impact parameter projection for $\chi\rightarrow3j$]{\includegraphics[width=0.34\textwidth]{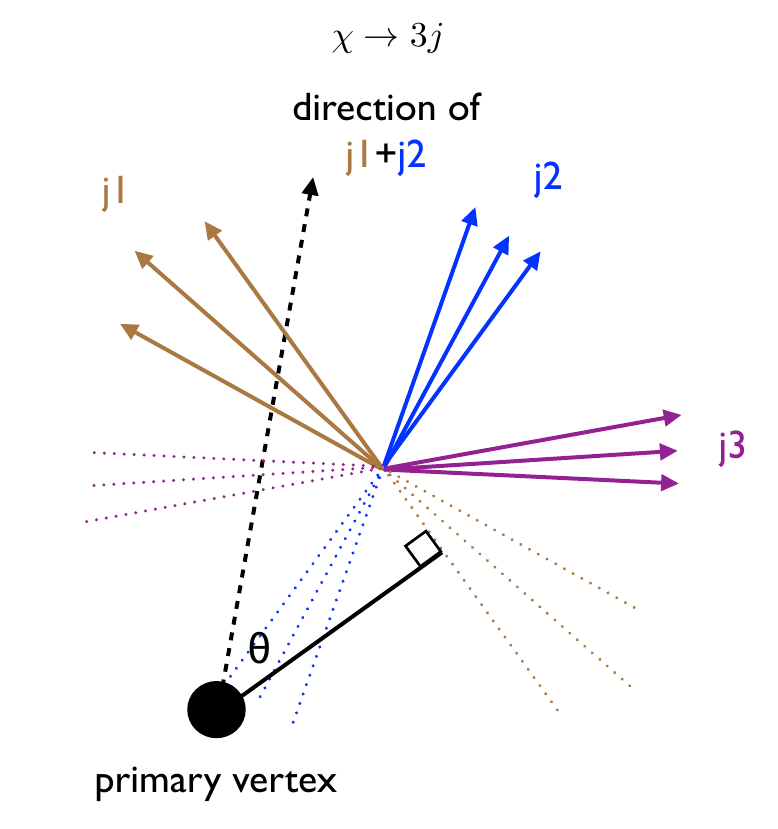}}
\caption{Schematic diagram showing the projection of the transverse impact parameter onto the dijet momentum vector. They point in the same direction if $|\theta| < \pi/2$.}
\label{fig:IP_schematic}
\end{center}
\end{figure*}

A multivariate discriminant is then formed from observables related to the tracks:%
\benum
\item The number of tracks at the displaced vertex;
\item The number of tracks in the largest cluster;
\item The root-mean-square (RMS) $L_{xy}^{\rm track}$ for each particle in a cluster, divided by the vertex $L_{xy}$;
\item The fraction of vertex tracks whose transverse impact parameter points in the same direction as the dijet momentum vector (i.e.~the projection of the transverse impact parameter onto the dijet momentum is positive; see Fig.~\ref{fig:IP_schematic}).
\eenum
Distributions of these observables are shown in Fig.~\ref{fig:multivariables}. For each event, all four observables are constructed from the displaced vertex and associated tracks. The probability that each observable value would come from the  signal and background distributions, denoted as $p^i_{\rm S}$ and $p^i_{\rm B}$, are calculated for each observable $i$. Finally, the multivariable discriminant $p$ is defined as follows:
\be\label{eq:MVA}
p = \frac{\prod_{i=1}^{n_{\rm obs}} p_{\rm S}^i}{\prod_{i=1}^{n_{\rm obs}} p_{\rm S}^i + \prod_{i=1}^{n_{\rm obs}} p_{\rm B}^i}.
\ee
The discriminant peaks towards $1$ for signal events and zero for background. Note that the product $\prod_{i=1}^{n_{\rm obs}} p_{\rm S}^i$ is, by definition, always large for signal events, while if there is even a single observable $j$ for which the signal and background look different ($p_{\rm B}^j\ll p_{\rm S}^j$), and so the product $\prod_{i=1}^{n_{\rm obs}} p_{\rm B}^i\ll\prod_{i=1}^{n_{\rm obs}} p_{\rm S}^i$, and $p\approx1$ for signal. By contrast, $p_{\rm S}^j\ll p_{\rm B}^j$ for a background event, and $p\ll1$. {\bf Thus, the discriminant $p$ has strong discriminating power even when only one of the observables exhibits a strong difference between signal and background}, with signal peaked towards 1 and background peaked towards 0.  CMS looks for events with one or more displaced vertices satisfying all of these criteria.

One possible limitation of this analysis it that several of the above observables rely on the fact that the long-lived particle decays into dijets. If it decays instead to a different final state, such as three jets or two jets and missing energy, a displaced dijet pair will not typically reconstruct the long-lived particle's momentum. In this case, the tracks will not be efficiently clustered together, possibly reducing the sensitivity of the displaced dijet analysis (see Fig.~\ref{fig:cluster_schematic}). Because the observables are combined in a multivariable discriminant, however, the CMS analysis efficiently separates signal from background if one of the variables has good discriminating power. It turns out that this variable is the {\bf vertex multiplicity:}~backgrounds from mis-reconstructed vertices typically have small track multiplicities, while signals originating from real displaced decays typically have high track multiplicities. Therefore we expect the CMS dijet search may apply to signal final states with $>2$ jets. We discuss further details in Section \ref{8TeV_dijet}.

\subsubsection{ATLAS displaced muon+tracks analysis}
The ATLAS analysis \cite{ATLAS-CONF-2013-092} looks for long-lived particles decaying into a muon and other charged particles. The particular model probed in their analysis is a long-lived neutralino produced from heavy squark decay; the neutralino in turn has a displaced decay to a muon plus jets via $Qd^{\rm c}L$-type RPV couplings.  The dominant background comes from a combination of both electroweak processes (for the muon) and QCD multijet processes (for the tracks). For the latter, the DV can arise where there are random crossings/mis-reconstruction of tracks, and where there is an interaction of hadrons from the primary vertex with material in the detector. Backgrounds from $b-$ and $c-$hadrons are suppressed by requiring a transverse impact parameter of $>2$ mm on the muon and associated tracks (which is sufficient to exclude most $b$-hadron tracks), and requiring a large invariant mass among the tracks and a well-separated, hard muon; all of these are rare to obtain from a heavy-flavour decay, which gives low-mass displaced vertices and soft, collinear leptons. The cut on impact parameter limits the efficacy of the search for particles with characteristic transverse decay distance $\lesssim$ mm.

The event sample is obtained with a trigger requiring a hard, central muon candidate in the muon system, but does not require a corresponding track. The analysis also requires the presence of a displaced vertex with at least five tracks with invariant mass $>10$ GeV. The experiment is sensitive to vertex displacements in the range of few mm to 20 cm. The ATLAS analysis looks for events with one or more displaced vertices satisfying all of these criteria. The simultaneous requirement of a hard muon and a multiple-track, high-mass displaced vertex is sufficiently strict to eliminate essentially all backgrounds at 8 TeV even without requiring that the muon be associated with the displaced vertex. It is therefore expected that, by additionally requiring that the muon be close to the displaced vertex, this analysis is expected to remain background-free throughout all future runs of the LHC \cite{priv_corr1}. 

The requirement of five tracks at the displaced vertex restricts its utility in scenarios were fewer tracks are produced. For example, for long-lived particles decaying to three leptons, there are too few tracks for the muon+tracks analysis to apply, although some bound could be placed from the non-observation of events in bins of lower track multiplicity. The exception is if there is a hadronically decaying $\tau$ in the final state, which can give more tracks; however, taus themselves have a finite lifetime and there is a gap between the displaced vertex and the $\tau$ decay point,  degrading the DV reconstruction. Therefore, this analysis applies primarily to particles which decay to quarks/gluons and muons; other analyses which specifically target dilepton final states are effective for other event topologies \cite{Chatrchyan:2012jna}. 

 \subsection{Monte Carlo methods}\label{sec:MC}
In  Sections \ref{8TeV_limits} and \ref{13TeV_prospect}, we present the results of a Monte Carlo analysis to estimate the bounds of existing displaced vertex searches on simplified models of WIMP baryogenesis, as well as anticipate the reach of the 13 TeV runs of the LHC. Signal events were generated with \texttt{Madgraph 5} \cite{Alwall:2011uj} using variants of the \texttt{RPVMSSM} model defined with \texttt{FeynRules} \cite{Alloul:2013bka}. The cross section for the wino model is renormalized to the next-to-leading-order (NLO) value using \texttt{Prospino} \cite{Beenakker:1999xh}. Parton-level events were showered with \texttt{Pythia 8} \cite{Sjostrand:2007gs}; the position of the long-lived particle decay is generated randomly using the momentum and lifetime of the long-lived particle and set manually in the \texttt{Pythia} event record.  Fast detector simulators are not expected to accurately reproduce the actual ATLAS and CMS displaced vertex reconstruction efficiencies for signal. Instead, we reconstruct the displaced vertices at particle truth level by associating with the vertex all final-state particles that can be traced back to the long-lived particle decay. This is then adjusted by a correction factor as described below. No MC simulation is needed (or possible) for backgrounds, as the background estimates are given directly in the experimental papers, which we approximately scale to higher energies and luminosity as needed.

The LHC experiments do not perfectly reconstruct the displaced vertices. To account for detector effects, we correct the particle truth-level MC efficiencies using the values from the CMS (displaced jets) and ATLAS (displaced muon + jets) analyses. In doing so, we assume that the entire difference between the particle-level and actual efficiencies arises from detector effects in reconstructing the tracks forming the displaced vertex. The dominant source of track mis-reconstruction is due to the degraded resolution in reconstructing tracks that originate far from the primary vertex, where the components of the tracker are spaced farther apart. Therefore, the ratio of experimental efficiency, $\epsilon_{\rm exp}$, to particle-level efficiency, $\epsilon_{\rm particle}$ is predominantly a function of the transverse position of the displaced vertex, $L_{xy}$, and depends more weakly on other factors of the displaced decay, such as the track $p_{\rm T}$ \cite{priv_corr3}. Thus, for each experimental analysis, we simulate the signals considered by ATLAS and CMS, determine a correction factor which is the ratio of the experimentally measured efficiency to our MC particle-level efficiency, $\epsilon_{\rm exp}/\epsilon_{\rm particle}$ for a given mean $\langle L_{\rm xy}\rangle$, and then apply this to different signal kinematics with the same $\langle L_{\rm xy}\rangle$. This allows us to approximately recast the experimental analyses, and also to consider the effects when we modify the cuts for the 13 TeV LHC runs. As far as we are aware, this approach is the best theorists can do to perform such a simulation; the precise determination of experimental sensitivity and efficiencies must be performed by the LHC collaborations.

We validate this approach using the data in the ATLAS and CMS displaced vertex searches. For both the displaced dijet and displaced muon+tracks analyses, the experimental collaborations give the signal selection efficiency for a variety of different transverse decay lengths and characteristic momenta for the displaced objects. We first consider the CMS displaced dijet analysis. As described above, they consider a benchmark model with a singly produced, scalar Higgs field $H$ that promptly decays into two long-lived scalars $S$, each of which decays at a mean transverse distance $\langle L_{xy}\rangle$ to two quarks. To test our signal MC, we were able to reproduce the signal acceptance at parton level (as defined in the CMS note) to within 10\% for each mass benchmark, excluding vertex reconstruction. The ratios of total signal efficiency $\epsilon_{\rm CMS}/\epsilon_{\rm particle}$ including all vertex selection criteria are shown in Table \ref{tab:eff_dijet_compare} for various mass benchmarks. 
 \begin{table}
\begin{tabularx}{\textwidth}{ |C|C|C|C| }
  \hline
    $\langle L_{xy}\rangle$ (cm) & $H$ mass (GeV) & $S$ mass (GeV)  & $\epsilon_{\rm CMS}/\epsilon_{\rm particle}$ \\
   \hline
  $\approx 3$ &1000 & 350 & 0.581 \\
    & 400 & 150 & 0.546 \\
     &  200 & 50  & 0.261 \\
       \hline
  $\approx 30$ &1000 & 350  & 0.297 \\
   & 400 & 150  & 0.231 \\
   &  200 & 50  & 0.14 \\
    \hline
  $\approx 300$ & 1000 & 350 & 0.157\\
  & 400 & 150  & 0.0654 \\
  \hline
\end{tabularx}
\caption{Comparison of our MC particle-level simulation and actual experimental efficiencies from CMS for various benchmark points of the displaced dijet analysis \cite{CMS-PAS-EXO-12-038}.}\label{tab:eff_dijet_compare}
\end{table}

We do a similar validation for the ATLAS displaced muon+tracks analysis. In the ATLAS note, they considered a SUSY signal model, with squark ($\tilde{q}$) pair production, followed by prompt decay to a jet and neutralino ($\tilde{\chi}^0$) each. The neutralinos are long-lived and, at a mean transverse distance $\langle L_{xy}\rangle$, decay through the RPV $Qd^{\rm c}L$ superpotential operator to a muon and two quarks. In Table \ref{tab:eff_muon_compare}, we show the ratios of total signal efficiencies $\epsilon_{\rm ATLAS}/\epsilon_{\rm particle}$ for various squark-neutralino benchmarks.

 \begin{table}
\begin{tabularx}{\textwidth}{ |C|C|C|C| }
  \hline
    $\langle L_{xy}\rangle$ (cm) & $\tilde{q}$ mass (GeV) & $\tilde{\chi}^0$ mass (GeV)  & $\epsilon_{\rm ATLAS}/\epsilon_{\rm particle}$ \\
  \hline
   $\approx 0.3$ & 700 & 494 & 0.46 \\
   & 700 & 108  & 0.40 \\
   & 1000 & 108 & 0.48 \\
   \hline
  $\approx 3$ &700 & 494 & 0.23 \\
    & 700 & 108 & 0.11 \\
     &  1000 & 108  & 0.076 \\
     \hline
    $\approx 30$ &700 & 494 & 0.12 \\
    & 700 & 108 & 0.068 \\
     &  1000 & 108  & 0.050 \\
       \hline
\end{tabularx}
\caption{Comparison of our MC particle-level simulation and actual experimental efficiencies from ATLAS for various benchmark points of the displaced muon+tracks analysis \cite{ATLAS-CONF-2013-092}.}\label{tab:eff_muon_compare}
\end{table}

Looking at Tables \ref{tab:eff_dijet_compare}-\ref{tab:eff_muon_compare}, it is evident that the ratio of $\epsilon_{\rm actual}/\epsilon_{\rm particle}$ is not precisely a constant for fixed $\langle L_{xy}\rangle$. For example, when the long-lived particles are highly boosted and the tracks at the displaced vertex are collimated, this can suppress vertex reconstruction efficiency, as noted in the CMS analysis. This degradation of experimental performance from the MC particle truth-level is seen for both the ATLAS and CMS analyses. Similarly, the vertex reconstruction worsens at very large $\langle L_{xy}\rangle$, most likely in a way that depends sensitively on the precise decay location on an event-by-event basis. Nevertheless, the correction factors between MC particle-level and the ATLAS and CMS analyses differ \emph{at most} by a factor of 2-3 for the same $\langle L_{xy}\rangle$, and often less for spectra with similar boosts and $\langle L_{xy}\rangle\sim$ cm, which is remarkable agreement given the very different particle kinematics for each benchmark point and the wide range of efficiencies observed for different displacements. Furthermore, if we use our method to recast constraints for high-mass states, an uncertainty in the fiducial cross section of ${O}(2)$ corresponds to an uncertainty in the mass bound of $O(10\%)$, and so our approach still allows us to draw strong qualitative conclusions on the regions of model space excluded by current searches and accessible at future runs of the LHC. In our subsequent analyses, we define our correction factor using the mass benchmark(s) with the lowest boost, as we focus on signals from direct pair production of the long-lived states, which typically have a small boost.

\begin{figure}[t]
\centering
\includegraphics[scale=0.865]{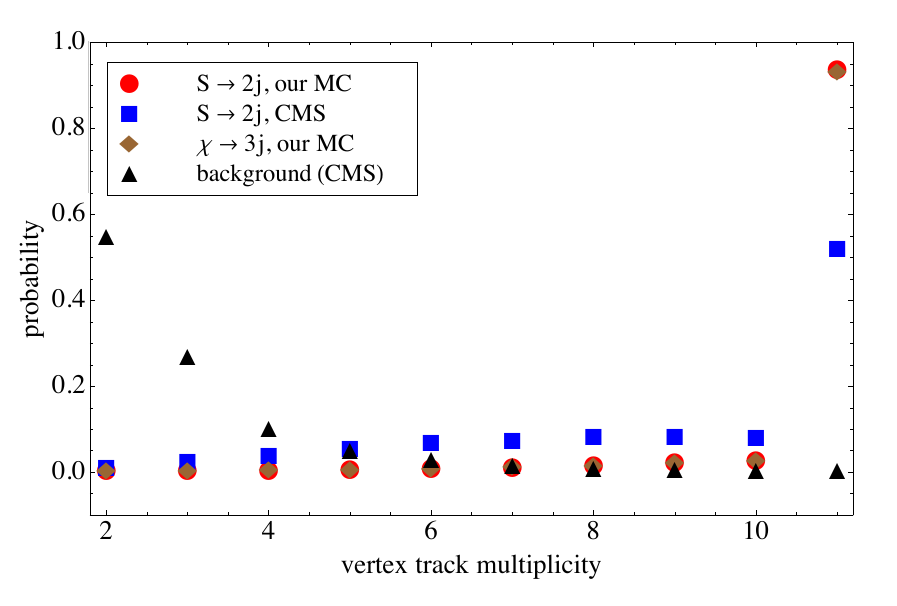}\hspace{0.4cm}\includegraphics[scale=0.85]{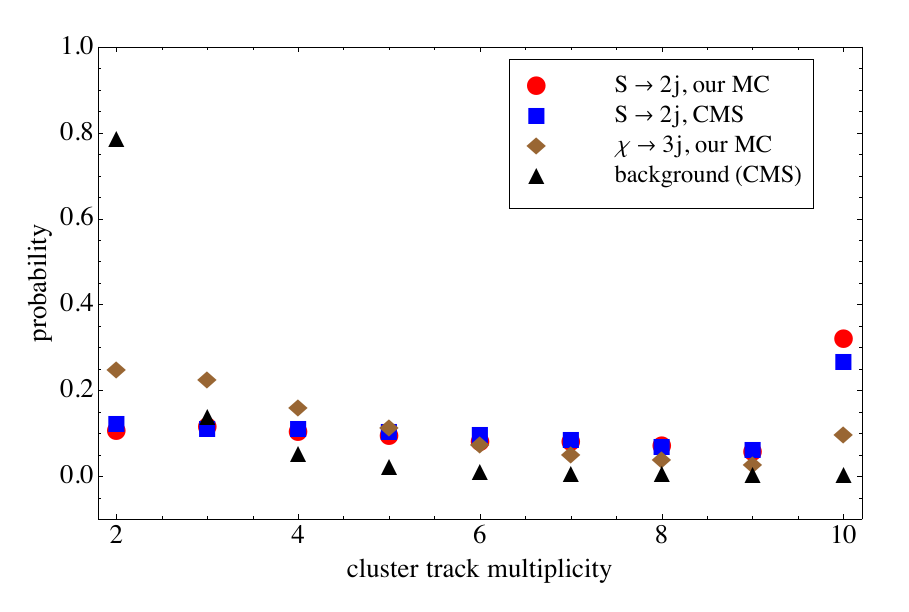}\\\vspace{0.7cm}
\includegraphics[scale=0.875]{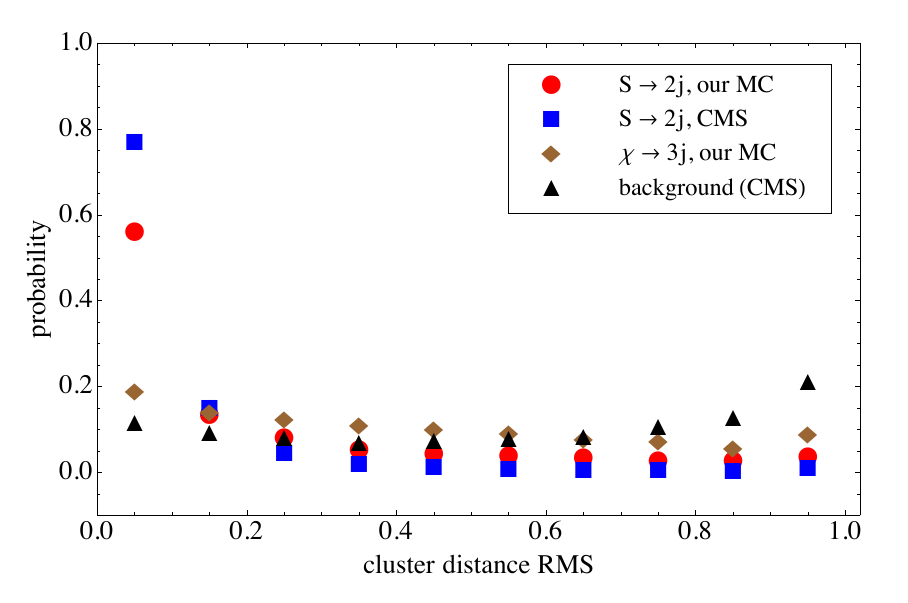}\hspace{0.4cm}\includegraphics[scale=0.85]{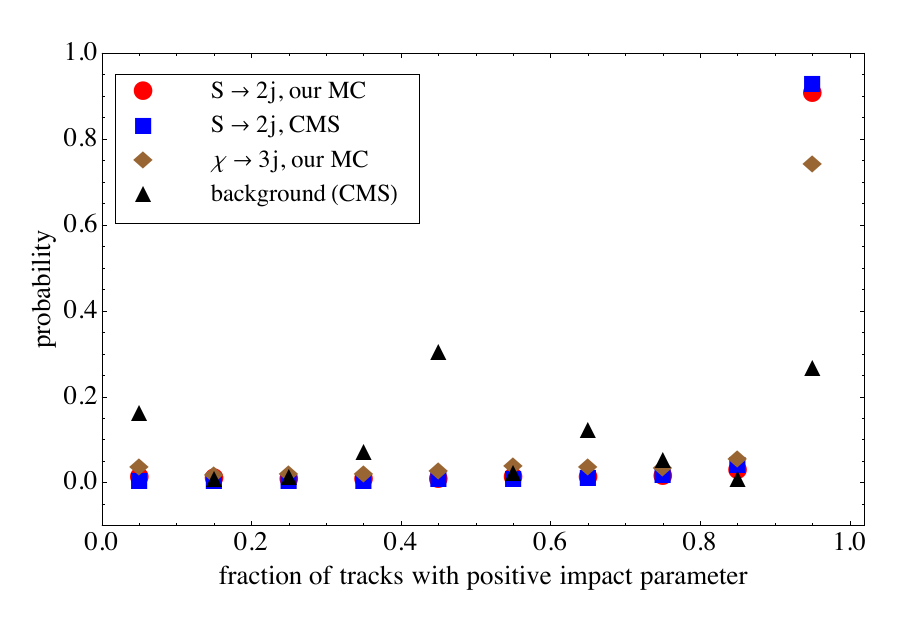}
\caption{Distributions for each of the four observables used in the multivariable discriminant $p$ of the CMS displaced dijet analysis \cite{CMS-PAS-EXO-12-038}. (Blue square) CMS result, $m_H=1$ TeV, $m_X=350$ GeV; (red circle) our particle-level simulation, $m_H=1$ TeV, $m_X=350$ GeV; (brown diamond) our particle-level simulation of wino pair production, with a displaced decay to three jets each ($m_\chi=500$ GeV); (black triangle) CMS backgrounds. For all signals, $\langle L_{xy}\rangle=29$ cm. }
\label{fig:multivariables}
\end{figure}

\begin{figure}[t]
\centering
\includegraphics[scale=1]{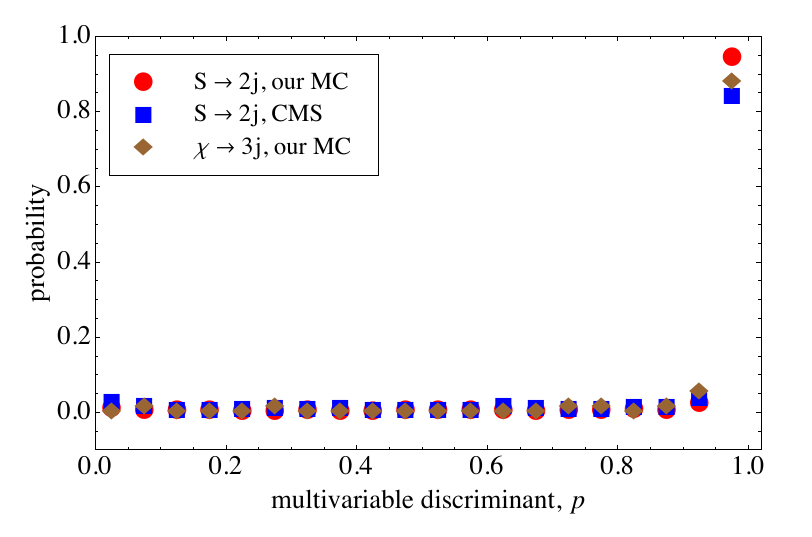}
\caption{Multivariable discriminant $p$ of the CMS displaced dijet analysis. The symbols are the same as Fig.~\ref{fig:multivariables}. For all plots, $\langle L_{xy}\rangle=29$ cm. The CMS note cuts on $p>0.8$ (0.9) for the ``high'' (``low'') selections of \cite{CMS-PAS-EXO-12-038}. }
\label{fig:discriminant}
\end{figure}

One may also assess whether our MC method is appropriate for reconstructing the multivariable discriminant $p$ used in the dijet analysis, Eq.~(\ref{eq:MVA}). To verify this, we use our simulation of the signal model considered in the CMS analysis and determine the distributions of the four observables in the discriminant:~the vertex track multiplicity, cluster track multiplicity, cluster RMS, and fraction of tracks with transverse impact parameter pointing in same direction as dijet momentum. We find that the track clustering algorithm using particle-level tracks most closely reproduces the CMS analysis when the clustering criterion is tightened to 1\% of the distance from the primary to secondary vertex. We show in Fig.~\ref{fig:multivariables} the distributions of the four observables, comparing the results of our particle-level MC simulation to the CMS results. While our simulations reasonably reproduce the cluster track multiplicity, cluster RMS, and impact parameter direction observables, it significantly overestimates the vertex track multiplicity; this is not surprising given that we are not including any detector effects from track reconstruction. However, we find that the multivariable observable is sufficiently robust such that these differences are relatively insignificant for the signal distribution of $p$. In Fig.~\ref{fig:discriminant}, we show the multivariable discriminant $p$, and we find that our simulation agrees with the observed CMS value to within 10\%\footnote{To determine the CMS value of the distribution of $p$, which is not shown in the paper, we generate many pseudo-events containing the four observables based on the CMS probability distributions in \ref{fig:multivariables} and assuming that the observables are uncorrelated; we then use Eq.~\ref{eq:MVA} to determine the $p$ distribution. Correlations will, if anything, tend to give larger values of $p$ for the CMS results, giving better agreement with our particle-level MC.}   For both 2- and 3-jet signals, the overwhelming majority of events is peaked at very high values of $p$, and therefore the signal efficiency of the CMS cut on $p$ is between 90-95\%. Therefore, our MC method adequately reproduces the multivariable discriminant to within an accuracy of 10\% for the purposes of recasting the CMS analysis.

\section{Limits from 8 TeV LHC}\label{8TeV_limits}
 \subsection{CMS displaced dijet analysis}
 \label{8TeV_dijet}
 
 We now study the bounds on the simplified models of Section \ref{sec:benchmarks}  from the CMS search for displaced vertices with dijets. The CMS search is relevant for all-hadronic final states originating from the displaced vertex, which is expected from $\chi$ decaying through operators such as $u^{\rm c}d^{\rm c}d^{\rm c}$ and $QQ(d^{\rm c})^\dagger$. A key difference between our models and the models directly constrained by CMS is that CMS targets decays into exactly two jets, whereas our final states have three or more jets. Thus,  one might na\"ively suspect that our signal models would not have a high efficiency of passing the CMS cut on these observables\cite{Graham:2014vya}. However, as we argued in the discussion below Eq.~(\ref{eq:MVA}), the multivariate discriminant remains effective for separating signal and background as long as one of its constituent observables has a very different distribution for signal and background. For the case of the CMS analysis, this observable is the vertex track multiplicity:~background displaced vertices are peaked towards small multiplicity, whereas signal tracks for both decays to two and three jets have track multiplicities $\gtrsim10$ (see Fig.~\ref{fig:multivariables}). Therefore, as we showed in the distribution of the multivariable discriminant $p$ in Fig.~\ref{fig:discriminant}, there is no reduction in discriminatory power for the 3-jet final state compared to the 2-jet final state, even though some of the observables in the multivariate discriminant are no longer powerful.  Therefore, one of our key results is that {\bf the CMS analysis strongly constrains final-state topologies beyond the intended displaced dijet analysis; in particular, long-lived particles decaying to three or more quarks are constrained.}

 \subsubsection{MSSM wino with RPV couplings}
In the RPV wino scenario, the new particles form an electroweak triplet, and charginos ($\tilde\chi^\pm$) and neutralinos ($\tilde\chi^0$) are produced through the SM electroweak interactions, and subsequently decay at a mean transverse distance $\langle L_{xy}\rangle$. As discussed in Section \ref{sec:benchmarks}, we assume that the charginos quickly decay into neutralinos, so that DVs arise only from neutralino decays. We impose all of the cuts of the CMS analysis, including an event-level $H_{\rm T}>300$ GeV requirement, two $p_{\rm T}>60$ GeV displaced jets at a vertex, and several other vertex selection criteria:~$m_{\rm track}>4$ GeV, $p_{\mathrm{T,\,vertex}}>8$ GeV, and the multivariate discriminant $p>0.9$ (0.8) for $\langle L_{xy}\rangle<30$ cm ($>30$ cm). We correct our particle-level Monte Carlo according to the prescription in Section \ref{sec:MC} in order to take  detector effects into account. Because the winos are pair-produced with low boost, we correct our MC at $\langle L_{xy}\rangle=3$, 30, and 300 cm using $\epsilon_{\rm CMS}/\epsilon_{\rm particle}$ averaged over the two lowest-boost mass benchmark points (namely, $m_H=1$ TeV, $m_S=350$ GeV; and $m_H=400$, $m_S=150$ GeV)\footnote{Using all three mass benchmarks at 3 and 30 cm changes the cross section limit by $\sim30\%$.}.

\begin{figure}[t]
\centering
\includegraphics[scale=0.9]{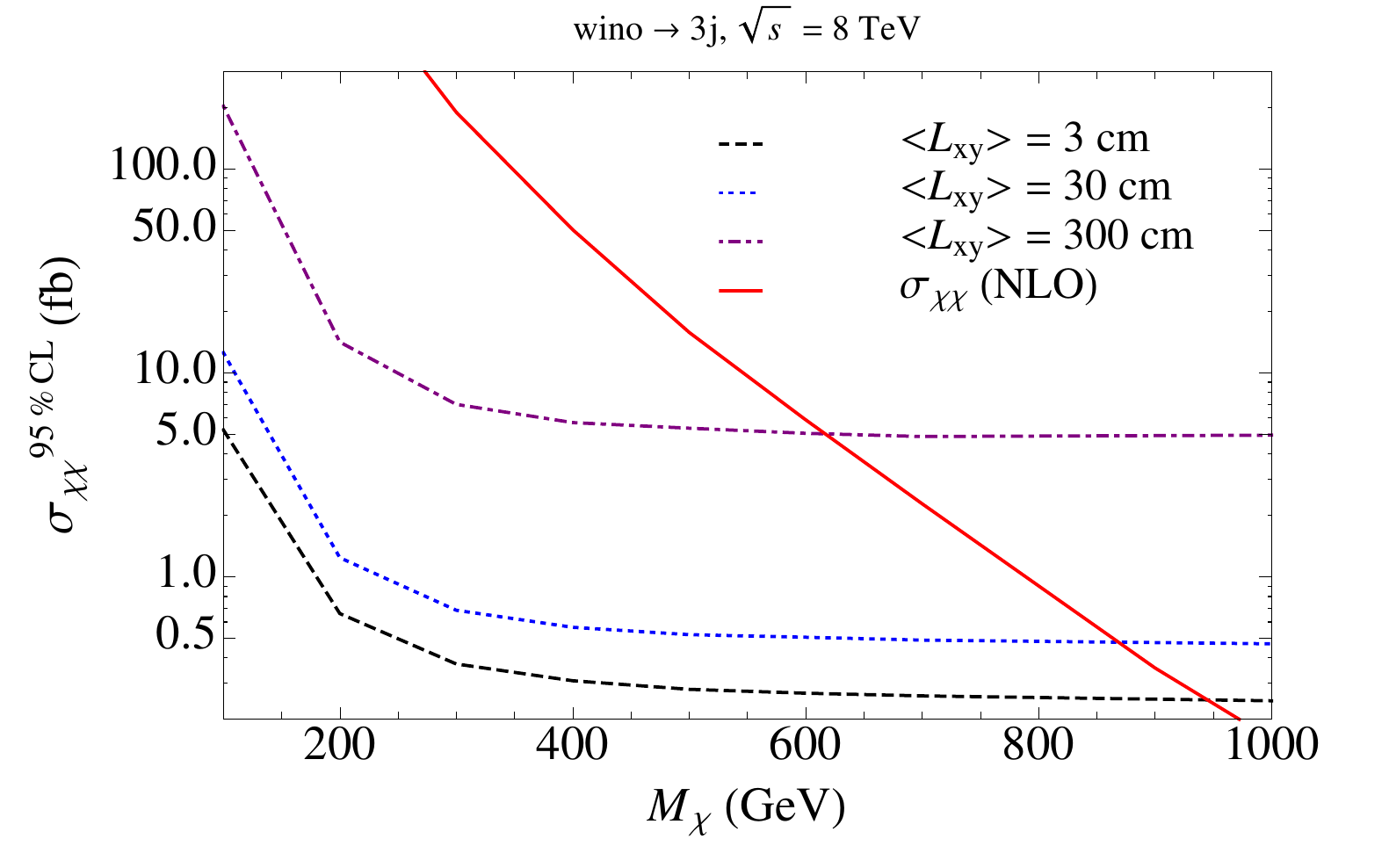}
\caption{Cross section limits for our simulation of {\bf winos decaying to three displaced jets} each at $\sqrt{s}=8$ TeV, with mean transverse vertex position $\langle L_{xy}\rangle$. The solid red curve shows the NLO production cross section. (The kinks in this and subsequent plots are due to the finite resolution of our parameter scan.)}\label{fig:wino_udd_8TeV_lowboost}
\end{figure}

This procedure allows us to approximately recast the CMS results for the hadronic RPV wino signal. We assume that the winos each decay into three light-flavour jets. We show our results for the bounds on the wino pair-production cross section as a function of mass in Fig.~\ref{fig:wino_udd_8TeV_lowboost}. For winos decaying in the most sensitive part of the detector ($\langle L_{xy}\rangle\approx3$ cm), winos are excluded up to about 950 GeV, while the bounds weaken to 625 GeV for very long lifetimes. Nevertheless, this excludes winos over much of the electroweak scale, and we see that the sensitivity of the CMS cuts is as high for displaced vertices with 3 jets as it is for 2-jet vertices. For decay lengths $\sim1$ mm, which is close to the lower lifetime bound from baryogenesis, the sensitivity to signal cross section should be a factor of a few worse than the 3 cm scenario, due to the presence of some displaced vertices that fail the impact parameter cut of 0.5 mm, and the reach becomes exponentially poorer for $\langle L_{xy}\rangle\ll 1$ mm. Therefore, the CMS analysis places strong bounds on WIMP baryogenesis through winos via hadronic RPV operators below the TeV scale over a wide range of relevant lifetimes.

 \subsubsection{Singlet coupled via Higgs portal}
We also consider the implications of the CMS search for the Higgs-portal singlet benchmark model, where each singlet fermion decays to three jets. Unlike the wino benchmark, which has a large production cross section and to which LHC8 is sensitive up to TeV-scale masses, the production cross section of the Higgs-portal singlet is much smaller due to the Higgs being off-shell in the production diagrams. We consider a benchmark point of $m_\chi = 150$ GeV, and the corresponding production cross section is
\bea\label{eq:Higgs_8TeV_xsec}
\sigma_{\chi\chi}^{\rm 8\,TeV}(m_\chi=150\,\,\mathrm{GeV}) &\approx& 0.23\,\lambda_{S\chi\chi}^2 \sin^2(2\alpha)\,\,\mathrm{fb},
\eea
where $\alpha$ is the Higgs-singlet scalar mixing angle. For mean transverse displacement $\langle L_{xy}\rangle$ of 3 cm, 30 cm, and 300 cm, we derive bounds 1.6 fb, 3.5 fb, and 57 fb, respectively. Therefore, even for maximal mixing between the singlet scalar and the SM-like Higgs, the CMS searches at 8 TeV do not impose any bound except for very large $\lambda_{S\chi\chi}$ and mixing. This motivates the development of search strategies to identify low-mass, long-lived particles at the 13 TeV runs of the LHC; we propose such strategies in Section \ref{13TeV_prospect}. Note that the cross section bounds on a 150 GeV singlet are  comparable to the cross section bounds on a 150 GeV wino in Fig.~\ref{fig:wino_udd_8TeV_lowboost}.

 \subsection{ATLAS displaced muon+tracks analysis}
 \subsubsection{MSSM wino with RPV couplings}
In this scenario, we consider an electroweak triplet wino decaying simultaneously into hadrons and leptons. This is the case when decay occurs through a $Qd^{\rm c}L$-type operator into a muon plus jets. Such a final state is probed by the ATLAS displaced muon+tracks analysis. We apply all of the ATLAS cuts, requiring a central, $p_{\rm T}>50$ GeV muon associated with a displaced vertex, as well as least four other tracks at the vertex with large transverse impact parameters with a combined $m_{\rm track}>10$ GeV for all tracks at the vertex. We correct our particle-level Monte Carlo according to the prescription in Section \ref{sec:MC}; because the winos are pair-produced with low boost, we correct our Monte Carlo using $\epsilon_{\rm ATLAS}/\epsilon_{\rm particle}$ for the lowest-boost benchmark point (namely $m_{\tilde q}=700$ GeV, $m_{\tilde\chi^0}=494$ GeV)\footnote{Using all three mass benchmarks changes the cross section limit by $\sim50\%$.}.

\begin{figure}[t]
\centering
\includegraphics[scale=0.9]{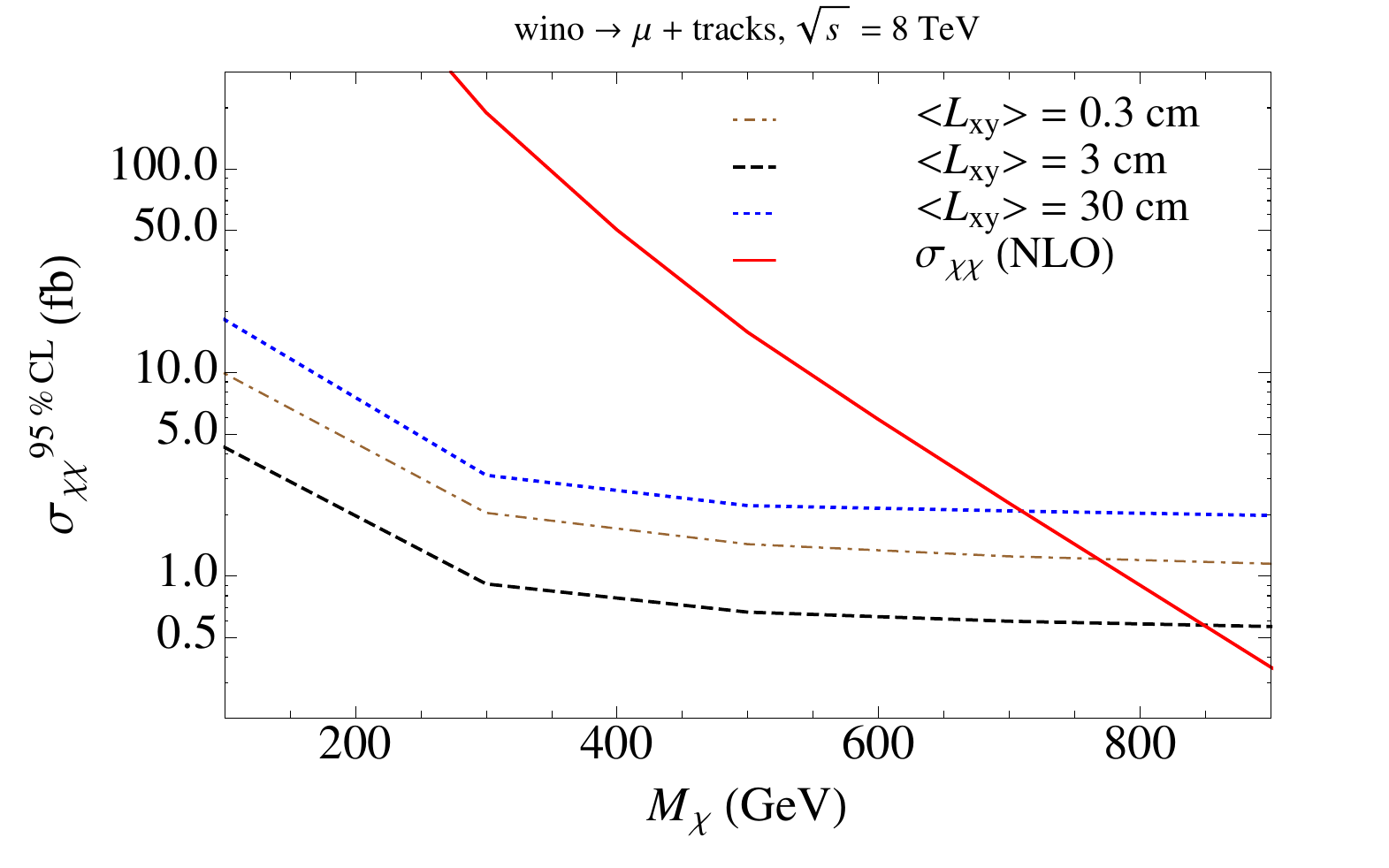}
\caption{Cross section limits for our simulation of {\bf winos decaying to one displaced muon+tracks} each, with mean transverse vertex position $\langle L_{xy}\rangle$. The solid red curve shows the NLO production cross section assuming 100\% branching fraction to muons.}
\label{fig:wino_muon_8TeV}
\end{figure}

Recasting the ATLAS results for the RPV wino signal, we show our results for the wino pair-production cross section bounds as a function of mass in Fig.~\ref{fig:wino_muon_8TeV}. The bounds range from 700-850 GeV for mean transverse displacements in the range of $0.3-30$ cm and 100\% branching fraction to muons, with the strongest bounds once again for $\langle L_{xy}\rangle\approx 3$ cm. This is comparable to the bound on hadronically decaying winos; this is reasonable, considering that both analyses are essentially background-free. For decay lengths $\sim1$ mm, which is close to the lower lifetime bound from baryogenesis, the sensitivity to signal cross section should be about an order of magnitude worse than the 3 mm scenario, due to the presence of some displaced vertices that fail the impact parameter cut of 2 mm, and the reach becomes exponentially poorer for $\langle L_{xy}\rangle\ll 1$ mm.  

 \subsubsection{Singlet coupled via Higgs portal}
Returning to the Higgs-portal singlet benchmark model, we recast the ATLAS muon+tracks search for the scenario where a pair of long-lived singlet fermions is produced through the Higgs portal, each of which decays to a muon and two quarks. As we found with the hadronic decay scenario, the cross section Eq.~(\ref{eq:Higgs_8TeV_xsec}) is $\lesssim0.1$ fb for $m_\chi=150$ GeV except at very large mixing and coupling, while we determine the bounds from the ATLAS analysis to be 6.7 fb, 2.4 fb, and 8.3 fb for $\langle L_{xy}\rangle=0.3$ cm, 3 cm, and 30 cm, respectively. Therefore, the ATLAS 8 TeV search does not place any significant constraints on baryogenesis-inspired singlets produced through the Higgs-portal if they decay to both baryons and leptons.

\section{Prospects at 13 TeV LHC}\label{13TeV_prospect}
We have seen that, despite the impressive sensitivity of the LHC DV searches at 8 TeV, there are limitations to their applicability to models of WIMP baryogenesis and other weak-scale baryogenesis theories. The LHC has strongly constrained the long-lived particle masses $\gtrsim500-1000$ GeV if they are charged under the electroweak interactions (or stronger forces). However, if the new particle is a gauge singlet, such as the off-shell Higgs-portal benchmark model, the LHC may have essentially no sensitivity, even if the new singlet has an $O(100)$ GeV mass.  The 13 TeV LHC running is where we can expect to probe this interesting regime.

The LHC 13 TeV run has two significant advantages over Run I:~it operates at a much higher centre-of-mass energy, allowing the production  of high-mass long-lived particles, and the higher luminosity enhances the sensitivity to lower-mass signals with small cross sections, such as the Higgs portal singlet benchmark model of WIMP baryogenesis. The downside is that trigger thresholds are higher at 13 TeV compared to Run I, and QCD background and pile-up rates are higher, potentially limiting the sensitivity of searches for new weak-scale long-lived particles. In this section, we estimate the reach of the 13 TeV LHC in constraining the benchmark models described in Section \ref{sec:benchmarks}. We also comment on how the search strategies can be modified to explore otherwise inaccessible parts of the parameter space.

\subsection{Sensitivity with current (single DV) searches}
\label{sec:1DV}

In this section, we consider minimal extensions of the existing CMS dijet and ATLAS muon+tracks searches at 13 TeV. Where necessary, we increase the event selection criteria in accordance with the higher trigger thresholds associated with higher energy and luminosity. Otherwise, we keep other aspects of the analysis the same. We now consider each search in turn.

\subsubsection{CMS displaced dijet analysis}
At 13 TeV, the displaced jet trigger used by CMS will have its $H_{\rm T}$ threshold increased from the current 300 GeV to 500 GeV \cite{CMS-TDR-012 }. This has negligible effect on long-lived particle masses well above 250 GeV, but suppresses the signal efficiency for lower-mass states. In doing our analysis, we keep all cuts the same as the 8 TeV analysis, but use the new trigger threshold. To estimate the background, we assume that the probability of random track crossing mis-reconstructing a DV scales linearly with pile-up, and consequently so does the background rate. The 8 TeV LHC running had an average number of primary vertices $\approx20$; we expect this to increase by an order of magnitude. We then multiply the 8 TeV background by a renormalizing factor to account for the suppressed QCD multijet rate of passing the $H_{\rm T}=500$ GeV at $\sqrt{s}=13$ TeV relative to the 300 GeV threshold at $\sqrt{s}=8$ TeV, while still applying a 60 GeV dijet cut associated with the high-level displaced trigger; this factor is approximately 0.75\footnote{We simulated this using a multijet sample with the matrix element matched to the parton shower using the $k_\perp$ scheme \cite{Alwall:2008qv} in \texttt{Madgraph} + \texttt{Pythia}.}. Since the original background estimate from the CMS analysis is 86 ab \cite{CMS-PAS-EXO-12-038}, accounting for the change in kinematics from 8 TeV to 13 TeV and an order-of-magnitude increase in pile-up results in a background rate of $\approx750$ ab at 13 TeV, implying an $O(1000)$ number of background events at high luminosity, in contrast with the nearly zero background situation in Run I. While there are several approximations that go into this estimate (for example, we do not include the scaling of track multiplicities from 8 to 13 TeV, as these are generally poorly modelled by MC, but will likely increase background rates), it provides a reasonable and conservative background estimate up to an $O(1)$ factor.

Using the new background estimate  and the same signal simulation as in Section \ref{8TeV_dijet}, we use a $\chi^2$ test \cite{Fox:2011pm} to determine the luminosity needed for a $2\sigma$ bound on our signal models. We require an estimate of the systematic uncertainties on the background. The ``low'' selection criteria of the 8 TeV analysis have an approximately 30\% systematic uncertainty \cite{CMS-PAS-EXO-12-038}, which is the set of cuts with the smallest systematic uncertainty in the CMS analysis; we consider both an optimistic projection, in which the systematic uncertainty is reduced to 10\% with the accumulation of more data and better understanding of the detector response, and a pessimistic projection in which the systematic uncertainty remains at 30\%.\\

\begin{figure}[t]
\centering
\subfigure[$2\sigma$ reach, 10\% systematic uncertainty]{\includegraphics[scale=0.54]{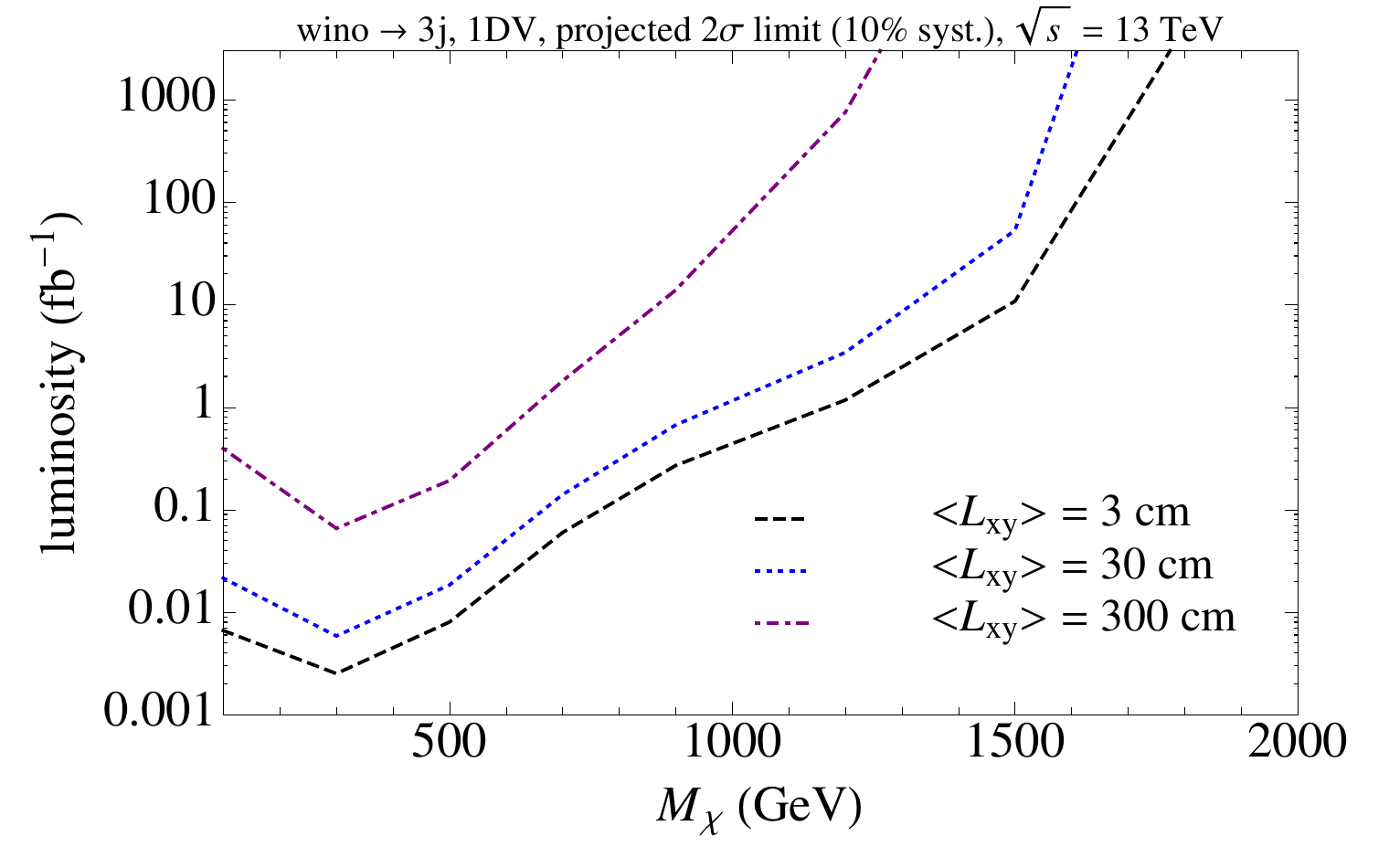}}
\subfigure[$2\sigma$ reach, 30\% systematic uncertainty]{\includegraphics[scale=0.54]{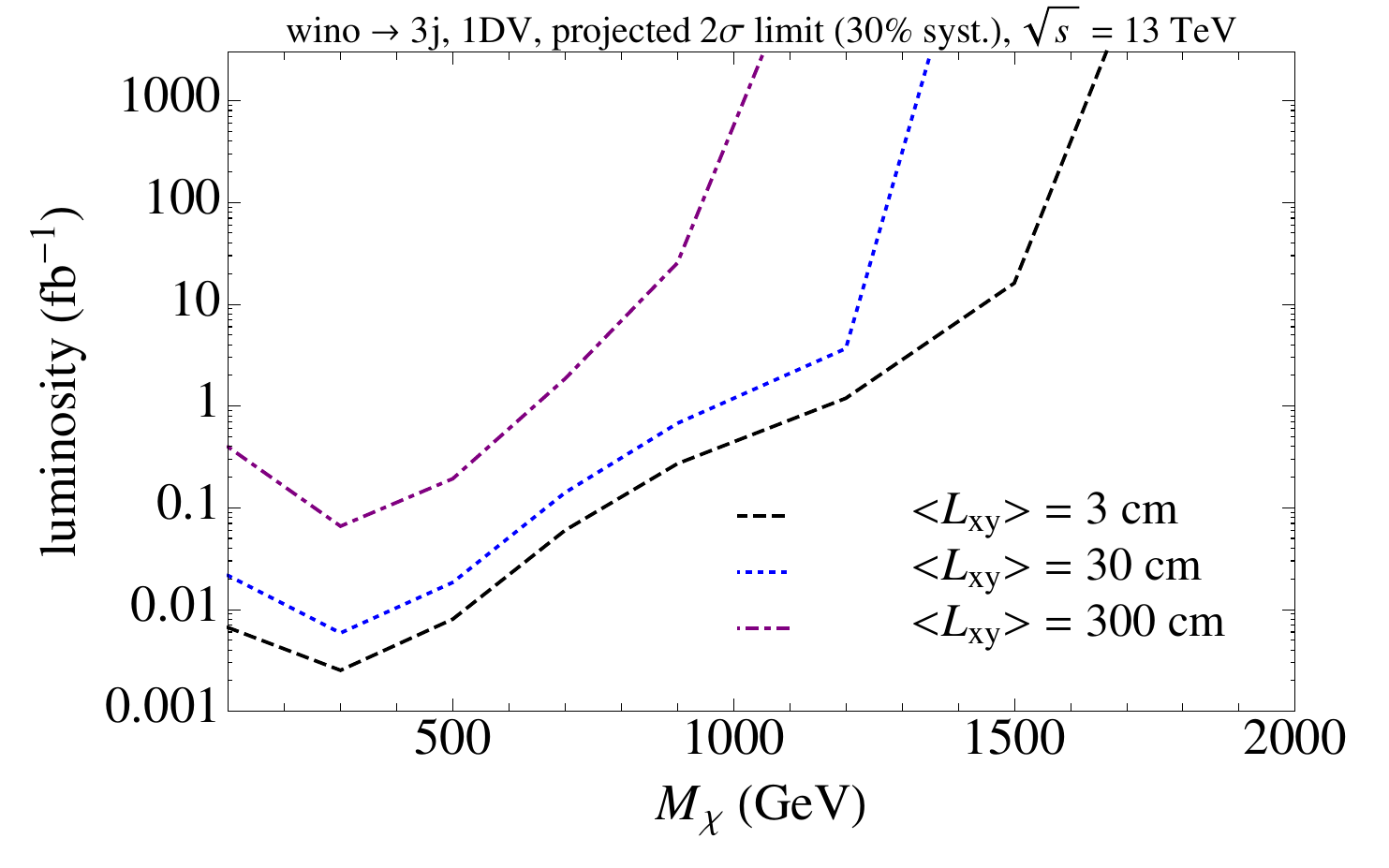}}
\caption{Luminosity required at 13 TeV for a $2\sigma$ limit on {\bf wino pair production decaying to three displaced jets} each, requiring  {\bf 1 DV tag}.}
\label{fig:wino_udd_13TeV_1DV_2sigma_lowboost}
\end{figure}

\noindent {\bf MSSM wino with RPV couplings:} 
  In Fig.~\ref{fig:wino_udd_13TeV_1DV_2sigma_lowboost}, we show the $2\sigma$ sensitivity on the wino benchmark model at 13 TeV. We see that, depending on $\langle L_{xy}\rangle$, lower bounds on the wino mass in the range of $1.3-1.8$ TeV are possible with a 10\% systematic uncertainty; if the current 30\% systematic uncertainty cannot be improved upon, the mass sensitivity drops by about $10-20$\%. For each curve, we see a rapid increase in the luminosity needed for $2\sigma$ sensitivity at very high $\chi$ mass; when the mass is too high, the signal-to-background ratio is too small for detection due to the dominance of systematic uncertainty. For this reason, we propose measures to reduce the background rate and consequently enhance $S/B$. In particular, we show in Section \ref{sec:2DV}  that requiring two DV tags results in an essentially background-free measurement, enhancing the sensitivity for very low signal rates/high wino masses. Alternatively, raising the $H_{\rm T}$ and $p_{\rm T}$ thresholds above the values necessary for the trigger could suppress backgrounds and improve the sensitivity at high masses, but our inability to simulate backgrounds hinders our ability to explore these possibilities.\\

\noindent {\bf Singlet coupled via Higgs Portal:} We repeat the analysis and project the 13 TeV sensitivity for the singlet meta-stable WIMP produced through the Higgs portal. Recall that we considered a benchmark point $m_\chi=150$ GeV and want to determine the LHC reach for the parameter $\lambda_{S\chi\chi}\sin2\alpha$, where $\alpha$ is the Higgs-singlet mixing angle. The cross section is:
\bea
\sigma_{\chi\chi}^{\rm 13\,TeV}(m_\chi = 150\,\,\mathrm{GeV})&\approx&0.93\,\lambda_{S\chi\chi}^2\sin^2(2\alpha)\,\,\mathrm{fb}.
\eea
Large values of the coupling and mixing angle ($\gtrsim1$) can be probed with $50\,\,\mathrm{fb}^{-1}$. A sharp fall-off of the sensitivity at smaller couplings and mixings occurs due to the signal rate becoming smaller than the systematic uncertainty. Smaller products of coupling and mixing angle can only be probed at high luminosity if the background can be suppressed using the same techniques as we propose for the wino model.

\begin{figure}[t]
\centering
\subfigure[$2\sigma$ reach, 10\% systematic uncertainty]{\includegraphics[scale=0.51]{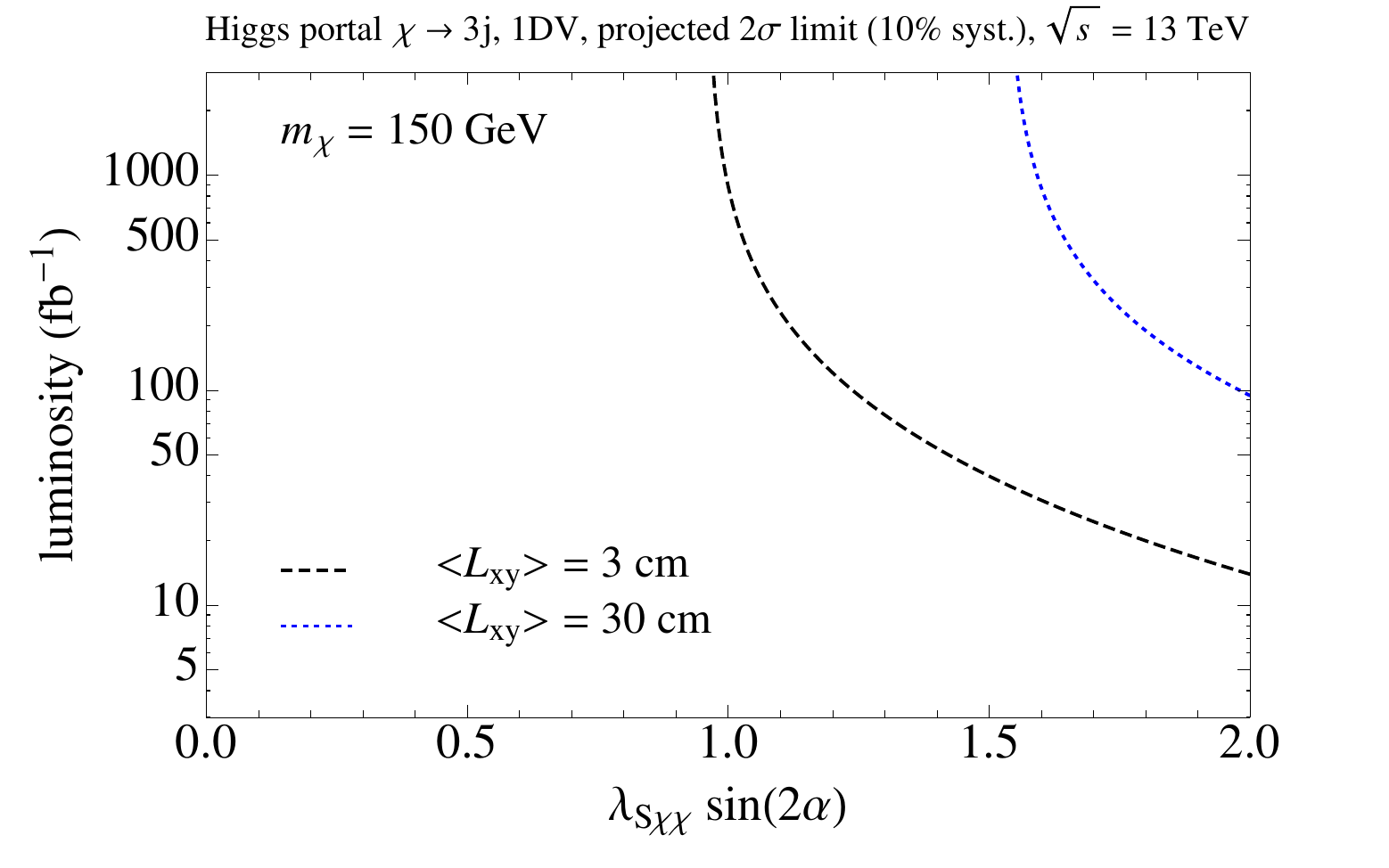}}
\subfigure[$2\sigma$ reach, 30\% systematic uncertainty]{\includegraphics[scale=0.51]{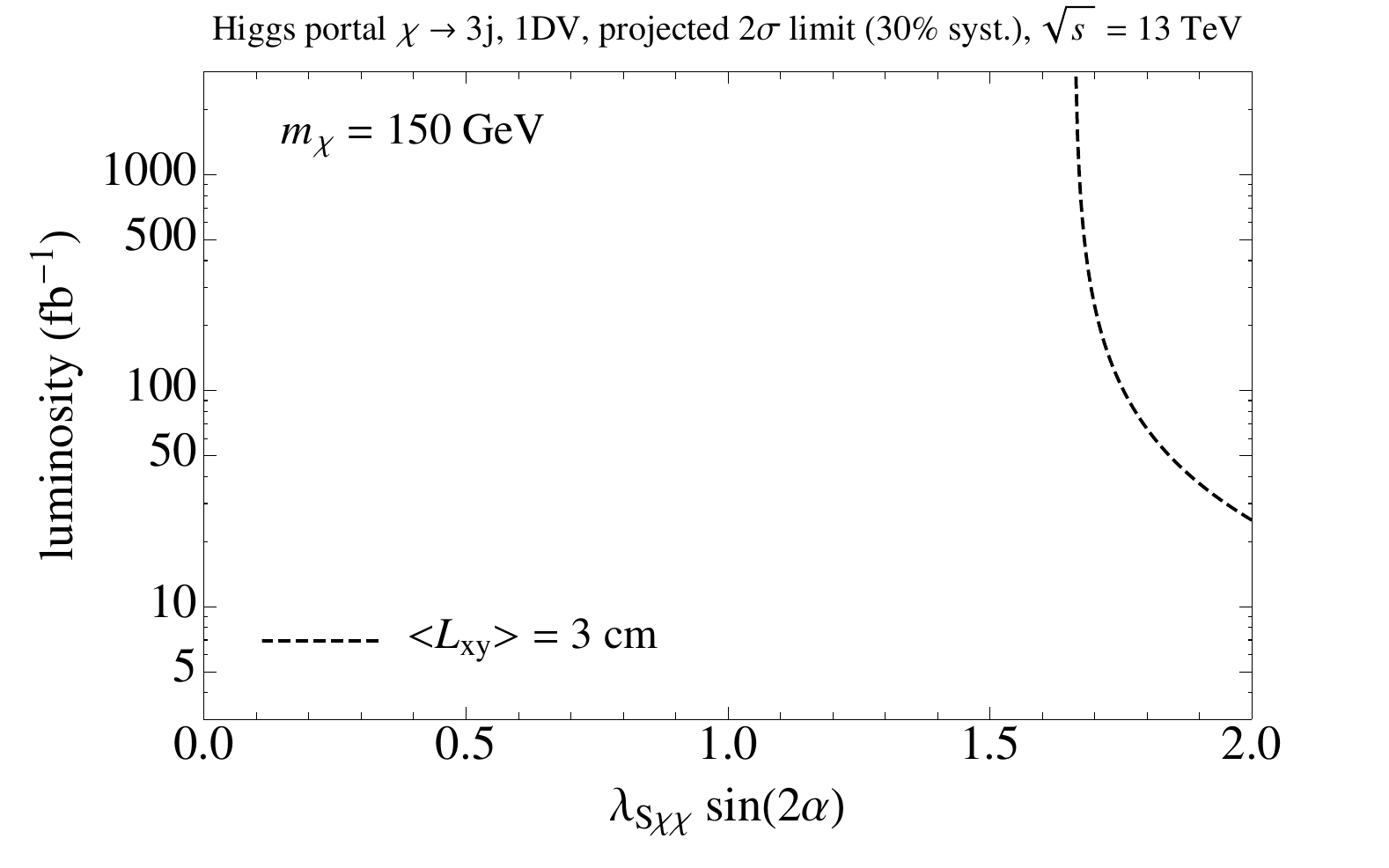}}
\caption{Luminosity required at 13 TeV for a $2\sigma$ limit on {\bf singlet pair production through the Higgs portal decaying to three displaced jets} each, requiring  {\bf 1 DV tag} ($m_\chi=150$ GeV). $\langle L_{xy}\rangle=300$ cm cannot be constrained.}
\label{fig:higgs_udd_13TeV_1DV_2sigma_lowboost}
\end{figure}

 \subsubsection{ATLAS displaced muon+tracks analysis}
 \label{sec:ATLAS_13}
 
The 8 TeV ATLAS analysis for DVs with a muon and tracks is essentially background free:~they expect $0.08\pm0.08$ events \emph{without} requiring that the muon be associated with the displaced vertex (they do impose this requirement for the signal). By imposing a muon and other displaced tracks to be at the same vertex, the analysis can remain background-free throughout the 13 TeV running of the LHC \cite{priv_corr1}. Furthermore, the 8 TeV analysis uses a 50 GeV muon trigger; this is not expected to rise substantially at 13 TeV\cite{priv_corr2}, so we keep all kinematic thresholds the same for the 13 TeV analysis.

Because the significance of any observed signal depends on the precise background estimate, which we cannot do in a realistic way, we instead take as our benchmark the observation of {\bf 3 signal events}. This is useful in a number of ways:~there is a 95\% probability that one or more events will actually be seen at the LHC, and any background estimate predicting $\leq0.01$ events allows for $\gtrsim5\sigma$ discovery with 3 signal events for Poisson statistics. Therefore, this is a reasonable benchmark to use for comparing to the $2\sigma$ reach of a search with large backgrounds, and we apply it to all analyses that are expected to be essentially free of backgrounds.\\

\begin{figure}[t]
\centering
\includegraphics[scale=0.7]{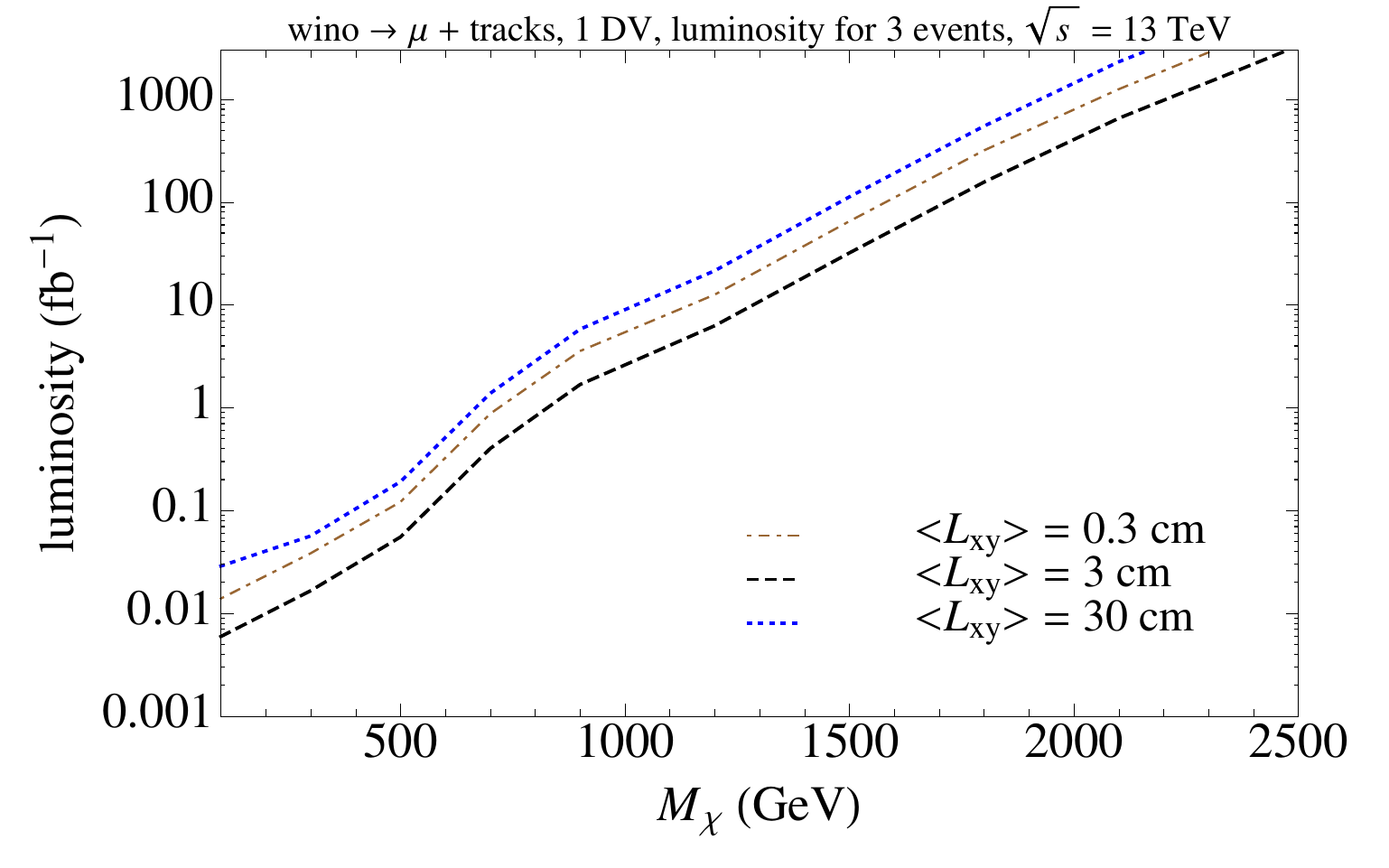}
\caption{Luminosity required at 13 TeV for three signal events for {\bf wino pair production decaying to a displaced muon+tracks} each, requiring {\bf 1 DV tag}. We use the three-event benchmark because the search is background-free and assume 100\% branching fraction to muons.}
\label{fig:wino_muon_13TeV_1DV_nobkd}
\end{figure}

\noindent {\bf MSSM wino with RPV couplings:} We show in Fig.~\ref{fig:wino_muon_13TeV_1DV_nobkd} the luminosity needed for three signal events at LHC13 as a function of the wino mass. We see that, depending on the lifetime, wino masses up to the range $2.1-2.5$ TeV can be probed. This is an extremely high mass reach for a weakly charged state, and it is entirely the result of the highly distinctive displaced nature of the final state, which allows for a background-free analysis. The main limiting factor for high mass winos is the efficiency of reconstructing the displaced vertices; one strategy for enhancing the sensitivity in this range is to  tighten kinematic cuts while relaxing some of the quality requirements on DV reconstruction, which may increase signal efficiency. Even if this is not done, however, a substantial range of the meta-stable WIMP baryogenesis parameter space with light winos will be probed at the LHC.\\

\begin{figure}[t]
\centering
\includegraphics[scale=0.7]{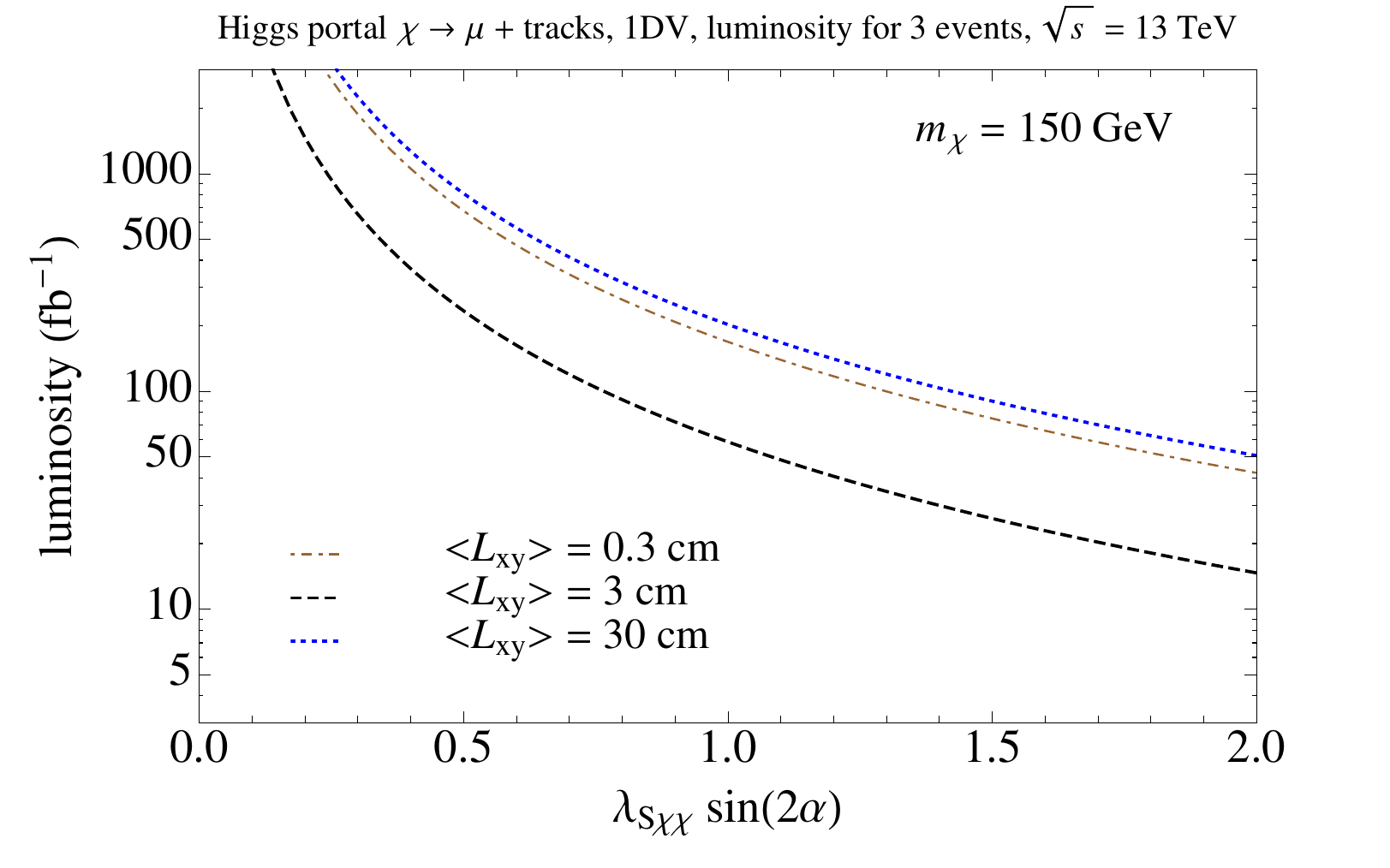}
\caption{Luminosity required at 13 TeV for three signal events for {\bf singlet pair production through the Higgs portal decaying to a displaced muon+tracks} each, requiring {\bf 1 DV tag} ($m_\chi=150$ GeV). We use the three-event benchmark because the search is background-free and assume 100\% branching fraction to muons. }
\label{fig:higgs_muon_highthreshold_lowboost}
\end{figure}

\noindent {\bf Singlet coupled via Higgs Portal:} We repeat the same analysis for the 13 TeV sensitivity to the production of singlets through the Higgs portal; the results are in Fig.~\ref{fig:higgs_muon_highthreshold_lowboost}. At high luminosities, the LHC is sensitive to relatively small couplings and mixing angle combinations of $O(0.1-0.3)$. Once again, the analysis at small coupling/mixing is limited by the signal efficiency; finding ways to improve signal efficiency while keeping a background-free experiment could further improve the prospects for detection.

\subsection{Improving sensitivity by double-vertex tagging}
\label{sec:2DV}
As we have seen, many representative WIMP baryogenesis models are constrained by the current data, or will be constrained by future running of the LHC by a straightforward application of the ATLAS and CMS DV searches. The displaced muon+tracks search is expected to continue to be background-free through the very high luminosity stages of the LHC, and so the current search  continues to remain sensitive. For the displaced dijet search, however, entirely hadronic backgrounds can be substantial at high luminosity, and models with very small signal rates are not within reach of the current search strategy.  This is especially true for the Higgs-portal singlet type of models, where signal events can be very rare for small Higgs-singlet mixing,  or when the production of $\chi$ proceeds through an off-shell Higgs. In this section, we demonstrate that requiring {\bf two DVs} completely removes any background without significantly reducing the signal, enhancing some of the search prospects for WIMP baryogenesis. Eliminating the possibility of backgrounds also allows the lowering of kinematic thresholds as much as possible to further enhance signal efficiencies, particularly for WIMP masses very close to the weak scale. 

The argument for requiring two DVs per event follows  from the symmetry structure of WIMP baryogenesis theories. The long lifetime of the particle(s) responsible for baryogenesis results from the tiny breaking of a $Z_2$ (or larger) symmetry. The production of $\chi$ is therefore suppressed unless it proceeds through interactions that respect the $Z_2$-symmetry; in other words, $\chi$ is always pair-produced at the LHC so that the final state is neutral under the stabilizing symmetry.  Consequently, the DVs naturally come in pairs as well; since it is exceedingly rare to find even a single fake DV in a background event (with cross sections in the $O(100)$ ab range), the probability of finding two is, for practical purposes, zero. For a more detailed estimate, see the Appendix.

According to the above arguments, the background rate should  remain zero for a 2 DV analysis even if we relax some of the selection cuts. The dijet DV search by CMS \cite{CMS-PAS-EXO-12-038} requires $H_{\rm T}>300$ GeV and only one DV with two $p_{\rm T}>60$ GeV jets, and this threshold will increase to $\sim500$ GeV in Run II. Signals with $m_\chi\lesssim150$ GeV, as in the Higgs portal models, have a  low efficiency of passing these cuts; consequently, these models are difficult to constrain even at high luminosity, as we showed in Section \ref{sec:1DV}. To improve the LHC sensitivity to low-mass long-lived particles, it would be ideal to relax the kinematic selection cuts. For example, we suggest exploiting the high jet-multiplicity for signal events using high-multiplicity triggers at Level-1 (in addition to inclusive $H_{\rm T}$ triggers). ATLAS  has plans for a level-1 4-jet trigger with $p_T>20$ GeV for the trigger-level regions of interest\cite{ATLAS-TDR-023, priv_corr2}, which correspond to jet-level thresholds of $\sim60-80$ GeV at level 1. The high-level trigger could then potentially capture these events using tracking information to reconstruct two displaced objects\cite{Buckley:2014ika}\footnote{For instance, the ATLAS FTK is expected to come online in 2016 and allow the use of tracking information on all events passing level-1; jets with many displaced vertices could be identified by the lack of tracks pointing to the PV.} Considering separately the $H_{\rm T}>500$ GeV vs.~4-jet, $p_{\rm T}>60$ GeV trigger selection, we find a comparable trigger+vertex selection efficiency for each. In our analysis below, we show results using the CMS $H_{\rm T}>500$ GeV displaced dijet trigger, but these could be improved by adding events passing other level-1 triggers such as the high-multiplicity triggers or triggers on associated objects, such as forward tagging jets in vector boson fusion (VBF) or leptons from Higgstrahlung production in the Higgs portal model\footnote{For the benchmark mass we use, $m_\chi=150$ GeV, the efficiency of passing the $H_{\rm T}>500$ GeV trigger is sufficiently high that using a VBF trigger is not expected to improve sensitivity, but for lower masses and consequently smaller signal $H_{\rm T}$, the Higgs associated triggers are very important.}. Also, the future development of techniques to provide tracking information at level 1 of the trigger can also yield substantial improvements. We emphasize that a 2 DV selection should be complementary to a 1 DV selection, which is useful in scenarios where the WIMPs are so long-lived that only one decays inside the detector volume; it should not be a problem to maintain both, since already both a 4-jet trigger and an $H_{\rm T}$ trigger are used together at level 1.

As an alternative to double DV tagging, it is possible to increase the signal-to-background ratio in the displaced dijet search by simply tightening the cut on the multivariable discriminant $p$, Eq.~(\ref{eq:MVA}). However, this is not particularly efficient. With MC simulations of the background $p$ distribution (from the CMS plots of the background distributions of each constituent observable and assuming they are uncorrelated, see Fig.~\ref{fig:multivariables}), we find that severely tightening the cut from $p=0.8$ to $p=0.99$ only reduces the background rate by a factor of $5$; this has a minimal reduction on the signal efficiency in our simulations, but a more substantial reduction is expected when detector effects are included. Regardless, this is not sufficient to ensure background-free events at high luminosities $\sim1000\,\,\mathrm{fb}^{-1}$. Such a cut is also very sensitive to modelling of the signal acceptance for $p$. By contrast, a 2 DV tagging method completely removes all backgrounds with only an ${O}(1)$ decrease in signal efficiency. 

While the ATLAS displaced muon+tracks analysis is expected to remain background-free at 13 TeV, the 2 DV tagging method could also be applied to this analysis if the estimated background unexpectedly increases in future runs. Alternately, there may be a way to increase the overall signal efficiency by relaxing some of the kinematic and DV tagging requirements but requiring 2 DV tags. We leave this to be done by the experimental collaborations, as it is beyond the scope of what is possible with our limited MC tools.

\begin{figure}[t]
\centering
\subfigure[$\langle L_{xy}\rangle=3$ cm]{\includegraphics[scale=0.43]{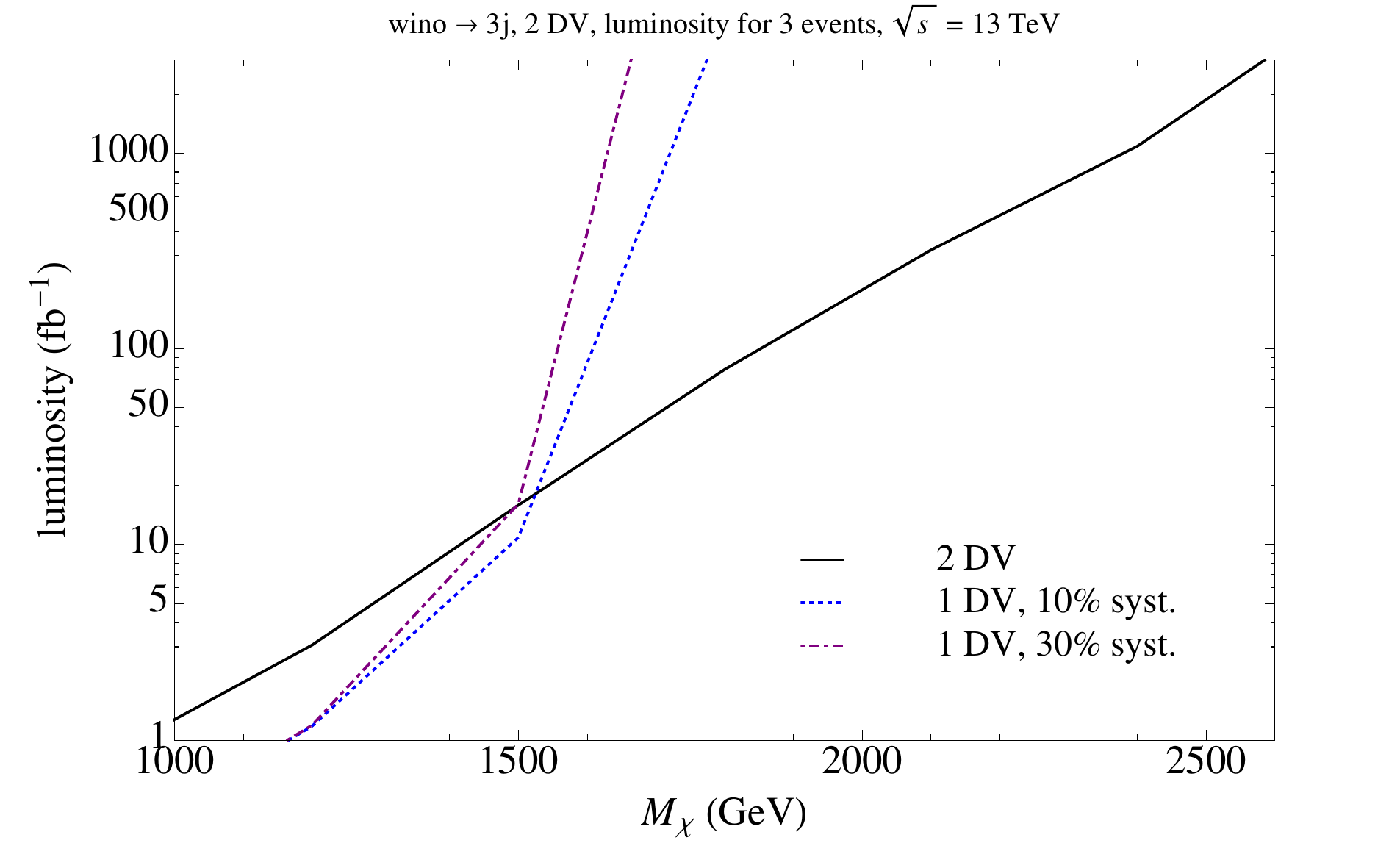}}
\subfigure[$\langle L_{xy}\rangle=30$ cm]{\includegraphics[scale=0.43]{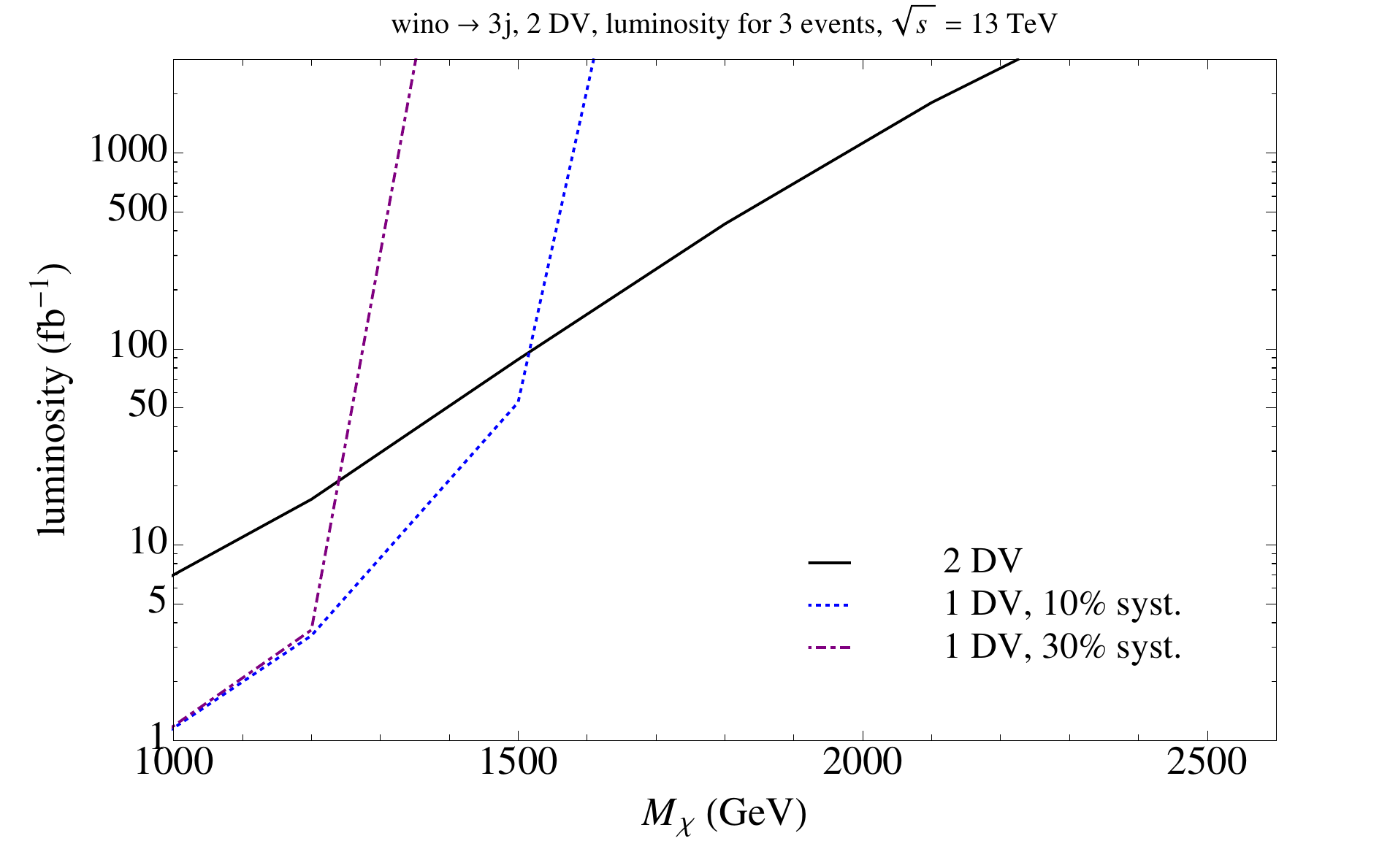}}
\caption{Luminosity required at 13 TeV for sensitivity to {\bf wino production decaying to three displaced jets} each, comparing our new three-event, background-free {\bf 2 DV tag} analysis to the $2\sigma$ reach of the  1 DV analysis (Fig.~\ref{fig:wino_udd_13TeV_1DV_2sigma_lowboost}). }
\label{fig:wino_udd_13TeV_2DV_3evt_lowboost}
\end{figure}

\begin{figure}[t]
\centering
\subfigure[$\langle L_{xy}\rangle=3$ cm]{\includegraphics[scale=0.44]{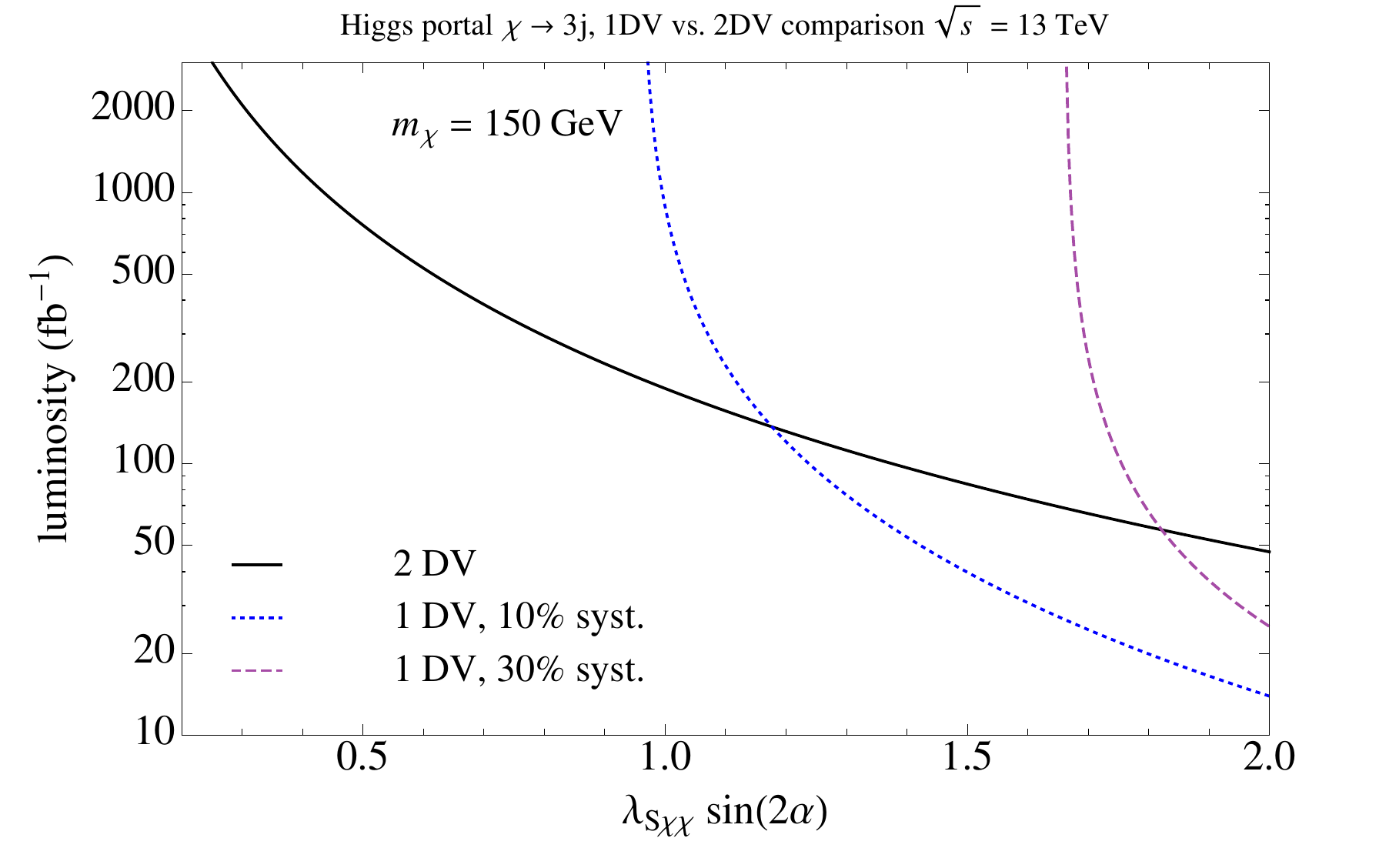}}
\subfigure[$\langle L_{xy}\rangle=30$ cm]{\includegraphics[scale=0.445]{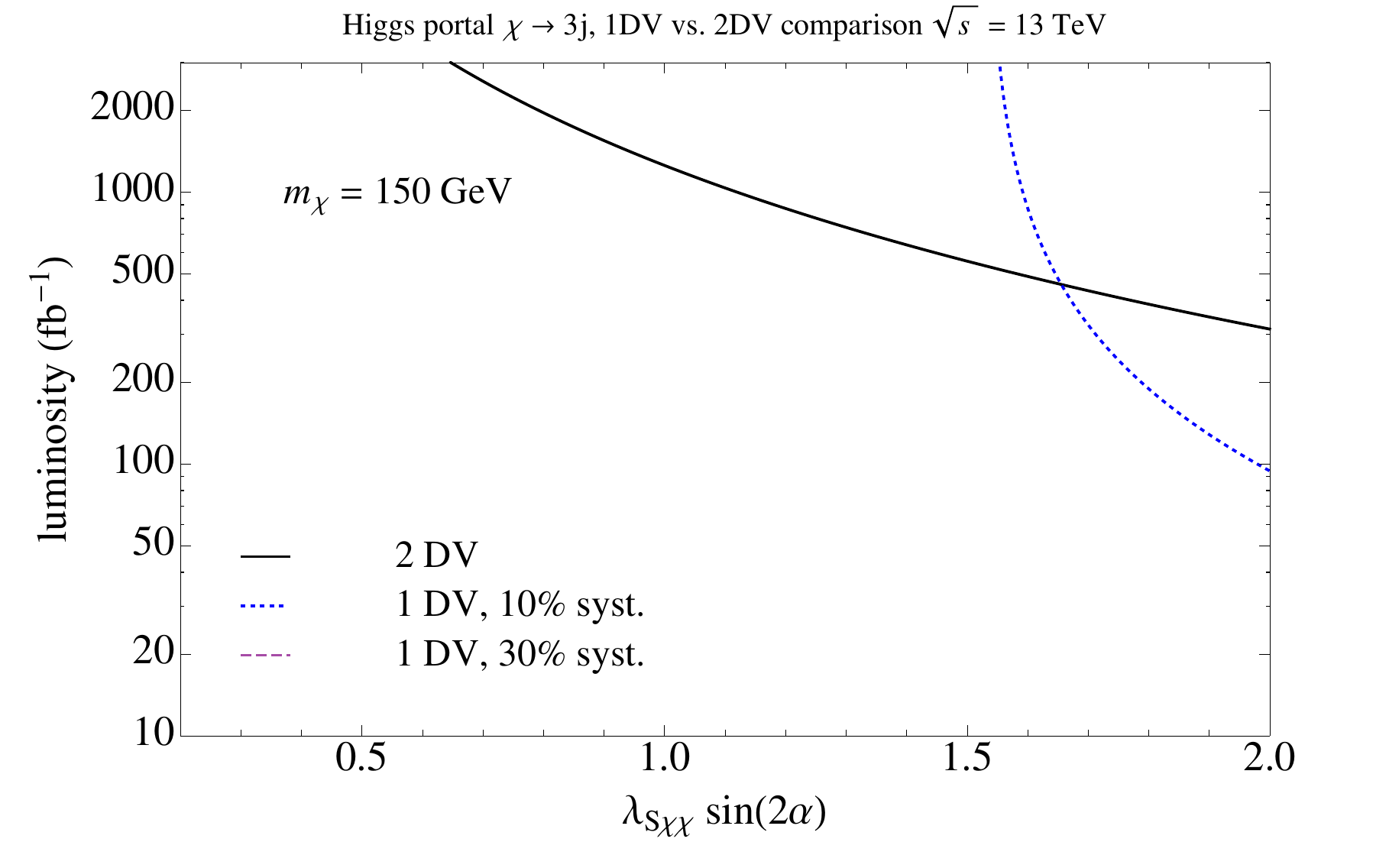}}
\caption{Luminosity required at 13 TeV for sensitivity to {\bf singlet pair production through the Higgs portal decaying to three displaced jets} each, comparing our new three-event, background-free {\bf 2 DV tag} analysis to the $2\sigma$ reach of the  1 DV analysis (Fig.~\ref{fig:higgs_udd_13TeV_1DV_2sigma_lowboost}). $\langle L_{xy}\rangle=300$ cm cannot be constrained.}
\label{fig:higgs_udd_highthreshold_lowboost}
\end{figure}

Among the two searches we are studying, backgrounds are only a concern for the displaced dijet analysis at 13 TeV. Therefore, we focus on this case to estimate the improvement obtained with 2 DV tagging using a higher-threshold version of the current CMS $H_{\rm T}$ trigger. We show our results in Fig.~\ref{fig:wino_udd_13TeV_2DV_3evt_lowboost} for the wino benchmark model,  and Fig.~\ref{fig:higgs_udd_highthreshold_lowboost} for the Higgs-portal singlet model, requiring three signal events as described in Section \ref{sec:ATLAS_13}. The true power of the 2 DV search emerges at high luminosity; at lower luminosities, the LHC is only sensitive to models with relatively large signal cross sections, and the importance of a background-free search is relatively less important.\\

\noindent {\bf MSSM wino with RPV couplings:} There is an approximately 40\% improvement in the mass reach for the 2 DV analysis compared to the 1 DV analyses, corresponding to an almost two order-of-magnitude increase in sensitivity to the production cross section. With 2 DV tags, winos with masses up to $2-2.6$ TeV are accessible at the LHC. Note that the trigger  is essentially irrelevant in this scenario, because the masses are sufficiently high that the signals are fully efficient for the planned CMS $H_{\rm T}>500$ GeV trigger. The double-DV tagging strategy is less effective at very large proper decay length, $L_{xy}\sim300$ cm, due to the fact that the DV reconstruction efficiency is smaller as the DV gets farther from the centre of the detector.\\

\noindent {\bf Singlet coupled via Higgs Portal:} The LHC with high luminosity is sensitive to mixing angle-product couplings of $O(0.2)$. This is a factor of $\sim4$ improvement in sensitivity to $\lambda_{S\chi\chi}\sin2\alpha$ for the 2 DV analysis compared to the 1 DV analysis at very high luminosity (see Fig.~\ref{fig:higgs_udd_highthreshold_lowboost}), and corresponds to a factor of $~20$ gain in the sensitivity to the production rate.

\section{Discussion and Conclusions}\label{concl}

In this work, we demonstrated that a large class of baryogenesis models based on the out-of-equilibrium decay of a new weak-scale particle, $\chi$, can be directly tested at the LHC by searching for events with displaced vertices. In these models, baryogenesis occurs in the weak-washout regime when the cosmic temperature drops below the particle mass, which optimally satisfies the  out-of-equilibrium condition for baryogenesis. The expected displaced decay length of $\gtrsim O(1)$ mm inside the collider automatically follows from the cosmological condition of weak washout. The long lifetime typically results from an approximate symmetry; while the symmetry-violating $\chi$ decay channels have small couplings, $\chi$ can be produced in pairs at the LHC through potentially sizeable symmetry-preserving interactions. The production of $\chi$ at the LHC is therefore exactly analogous to conventional dark matter searches at the LHC, although in this scenario, $\chi$ may not completely escape from the detector but instead decays to SM baryons and leptons at displaced vertices.

As a concrete, representative realization of the above scenario, we focused on the recently proposed WIMP baryogenesis mechanism, which can also address the DM-baryon abundance ``coincidence''. We organized the model space by considering a pair of simplified models which exemplify the possible $\chi$ production channels at the LHC. We estimated the limits and projected reach for these benchmark models when they decay entirely hadronically, or into a combination of hadrons and leptons, as constrained by two of the most sensitive searches at the LHC to-date:~the ATLAS displaced $\mu$+tracks and CMS displaced dijet searches. 

The methods we employ allow us to re-purpose existing LHC DV searches to the WIMP baryogenesis models. A key finding is that, although LHC searches target very specific final states, they are generally far more broadly applicable:~for example, we find that the CMS displaced dijet searches are also sensitive to 3-jet final states and other topologies with at least 2 jets. The methods we introduce can also be extended in a straightforward manner to other scenarios, such as purely leptonic decays, or decays to tops as motivated in the natural SUSY model of Ref.~\cite{Cui:2012jh}. There are other scenarios which require a departure from the minimal simplified models, such as if $\chi$ is produced from the decay of a heavy resonance as proposed in earlier theoretical studies \cite{Cui:2012jh}. 

The lower bound on the lifetime implied by the cosmic baryon abundance could also lead to decays in other parts of the detector, such as the muon chamber and calorimeter systems. Existing LHC  searches for long-lived particle decays in the muon chamber \cite{ATLAS:2012av} and the HCAL \cite{ATLAS-CONF-2014-041} currently give weaker limits ($\sim0.1-1\,\,\mathrm{pb}$) than the limits we find for decays in the most sensitive part of the tracker ($\sim1-10\,\,\mathrm{fb}$) due to lower signal efficiency due to less accurate reconstruction of displaced vertices in these systems, as well as large backgrounds for the HCAL search. This comes with a caveat:~the HCAL and muon system searches are derived from models of long-lived particles produced from Higgs decays, and often target the low mass long-lived particle region ($<m_H$), whereas the tracker searches are for heavier masses. Indeed, the lowest-mass CMS search (for a 200 GeV Higgs decaying into two 50 GeV long-lived scalars) sets the worst bound of 0.1 pb due to the low trigger efficiency. Extending the HCAL and muon system studies to higher mass (high $H_T$) region would allow for easier comparison with the tracker decay studies and facilitate projections to Runs II+. While a detailed consideration of the reach of the HCAL and muon system searches requires a dedicated analysis and is beyond the scope of this paper, the above considerations allow us to estimate that cross section limits from the HCAL/muon system searches will likely be an order of magnitude or two worse than the limits in the tracker system, but still probably sensitive to 1-10 fb cross sections with 3 $\mathrm{ab}^{-1}$ for $c\tau\sim$ few metres.

Our studies complement existing theoretical motivations from gauge- and anomaly-mediated SUSY, hidden valley models, twin Higgs theories, displaced SUSY, and others to provide a strong cosmological motivation for further developing dedicated displaced triggers (to complement existing ones), and generalizing the analyses of displaced vertex  of searches at the LHC. We also propose the tagging of two displaced vertices in inner-tracker DV searches, which generally applies to theories where the long-lived particles result from an approximate $Z_2$ symmetry, and completely eliminates backgrounds in the final states we study while allowing for the relaxation of the individual DV selection criteria to improve signal efficiency. We show that this approach may be able to improve the discovery potential  of signal cross sections by up to two orders of magnitude. Combining the current search strategies and our new proposal, we found that the high luminosity LHC at 13 TeV run can probe heavy electroweak charged long-lived WIMPs such as winos up to $2-2.5$ TeV masses for lifetimes $c\tau\sim1$ cm. By contrast, light displaced particles with masses $O(100)$ GeV, which are otherwise challenging to discover at LHC13, can be discovered with production cross sections as low as $O(50)$ ab for our benchmark models. These far exceed the reaches in search channels based on missing energy or prompt decays, due to the merit of low backgrounds in the displaced vertex channels.

The origin of the baryon asymmetry in our universe is an important unresolved question in particle physics and cosmology. The possibility that a large class of weak-scale baryogenesis models may be directly probed at the LHC in the displaced vertex channel presents an opportunity as exciting as missing energy searches for WIMP dark matter at the LHC, and should be vigorously pursued. We hope that our work will motivate further studies into optimizing and implementing improved displaced vertex searches at the LHC, leading to the possible discovery of the new physics responsible for baryogenesis in the coming years.

\section*{Acknowledgements}
We thank J.~Alimena, N.~Barlow, J.~Bouffard, J.~Incandela, E.~Izaguirre, V.~Jain, C.~Leonidopoulos, D.~Miller, P.~Saraswat, M.~Strassler, and R.~Sundrum for helpful discussions. We thank the organizers of the Particlegenesis workshop at KITP, during which a portion of this work was completed. YC is supported in part by NSF grant PHY-0968854 and by the Maryland Center for Fundamental Physics. This research was  supported in part by the National Science Foundation under Grant No.~PHY11-25915T, and by Perimeter Institute for Theoretical Physics. Research at Perimeter Institute is supported by the Government of Canada through Industry Canada and by the Province of Ontario through the Ministry of Research and Innovation.

\appendix

\section*{Appendix:~Estimate of 2 DV Background for Dijet Search}
\label{sec:appendix}
To confirm the zero-background hypothesis, we review the major sources of fake displaced vertices \cite{ATLAS-CONF-2013-092,Halyo:2013yfa}.
\benum
\item Vertices from real hadronic interactions with gas molecules outside the beampipe. Such vertices usually have low masses below the selection thresholds, but some tail of the distribution can fake the signal.
\item Purely random combinations of real or fake tracks. This is most important closer to the beampipe, where track densities are the highest.
\item Backgrounds from collimated $g\rightarrow b\bar b$ with displaced $b$ decays (where the $b$ happen to decay close together in the tracker)
\eenum
All of these backgrounds are sourced by QCD events. For a rough estimate of the background scaling, we simulate LO QCD dijet events with \texttt{Madgraph 5}, imposing generator-level cuts $H_{\rm T}>325$ GeV, $p_{\mathrm{T},j}>120$ GeV (assuming that the jet splits into two 60 GeV jets per DV) and $|\eta_j|<2$, and calculate a probability for obtaining a DV passing cuts from a given jet. The cross section is $\sigma_0\simeq3\times10^4$ pb at 8 TeV. The number of background DVs  after all CMS selection cuts (including the above basic kinematic cuts and the displaced vertex selection cuts) is $\sigma_{\rm 1\,DV}\approx0.1$ fb \cite{CMS-PAS-EXO-12-038}. Therefore a crude estimate of the probability that one of the QCD events gives a displaced vertex passing all cuts is:
\be
 \xi_{\rm DV}\equiv \frac{\sigma_{\rm 1DV}}{\sigma_0}\sim3\times10^{-9}.
 \ee
 Assuming that the reconstruction of background DVs are uncorrelated \cite{Halyo:2013yfa}, we require 2DVs in the event instead while keeping the same DV selection cuts for each of the DV, finding $\sigma_{\rm 2 DV}\sim\sigma_{\rm 1 DV}\xi_{\rm DV}\sim3\times10^{-10}$ fb for Run I data. In subsequent runs, the multijet rate is larger, and pileup contributions to accidental crossings is also more significant; each of these effects is an $O(1-10)$ factor. However, the background estimate is so low that, even with these enhancing factors and accounting for potentially large errors in our estimate, this is practically zero even at $\sqrt{s}=13$ TeV and large integrated luminosity. This estimate should be confirmed by the experimental collaborations.
\bibliographystyle{JHEP}
\bibliography{WIMPbg_ref}

\end{document}